\documentclass[review]{elsarticle} 
\makeatletter
\def\ps@pprintTitle{%
 \let\@oddhead\@empty
 \let\@evenhead\@empty
 \def\@oddfoot{\centerline{\thepage}}%
 \let\@evenfoot\@oddfoot}
\makeatother

\usepackage{lineno,hyperref}

\usepackage[labelsep=period,labelfont=bf]{caption} 
\usepackage{amsmath}

\usepackage{hyperref}
\usepackage[hyperref]{xcolor}
\hypersetup{
  colorlinks=true,
}

\begin{document}

\begin{frontmatter}

\title{Sound transmission of periodic composite structure lined with porous core: rib-stiffened double panel case}


\author[mymainaddress]{Hou Qiao}

\author[mymainaddress,mysecondaryaddress]{Zeng He\corref{mycorrespondingauthor}}
\cortext[mycorrespondingauthor]{Corresponding author}
\ead{hz@mail.hust.edu.cn}

\author[mymainaddress,mysecondaryaddress]{Wen Jiang}

\author[mythirdaddress]{Weicai Peng}

\address[mymainaddress]{Department of Mechanics, Huazhong University of Science $\&$Technology, Wuhan, China}
\address[mysecondaryaddress]{Hubei Key Laboratory for Engineering Structural Analysis and Safety Assessment, Huazhong University of Science $\&$ Technology, Wuhan, China}
\address[mythirdaddress]{National Key Laboratory on Ship Vibration and Noise, China Ship Development and Design Center, Wuhan, China}

\begin{abstract}


Porous materials are effective for the isolation of sound with medium to high frequencies, while periodic structures are promising for low to medium frequencies. 
In the present work, we study the sound insulation of a periodically rib-stiffened double-panel with porous lining to reveal the effect of combining the two characters above. 
The theoretical development of the periodic composite structure, which is based on the space harmonic series and Biot theory, is included.
The system equations are subsequently solved numerically by employing a precondition method with a truncation procedure.
This theoretical and numerical framework is validated with results from both theoretical and finite element methods.
The parameter study indicates that the presence of ribs can lower the overall sound insulation, although a direct transfer path is absent.
Despite the unexpected model results, the method proposed here, which combines poroelastic modeling and periodic structures semi-analytically, can be promising in broadband sound modulation.

\end{abstract}


\end{frontmatter}


\section{Introduction}

Owing to their high stiffness-to-weight ratio, multipanel structures are widely used in engineering applications, such as aircrafts, underwater, and architectural structures.
Their acoustic performance has been studied for a long time \cite{Sharp1978NCE,Lin1977TJotASoA,Hongisto2006AAuwA}. 

Composite multipanel structures without any attachments or fillings are always the simplest to operate. Both theoretical, and experimental and numerical methods are developed with regard to their sound transmission loss (STL); for example, the theoretical models by Xin \cite{Xin2009AJ}, Sakagami \cite{Sakagami2002AA} (with experiments) and the semi-empirical models by Sharp \cite{Sharp1978NCE}, Gu \cite{Gu1983CJoA}, Davy \cite{Davy2009BA}. These prediction models were reviewed and compared by Hongisto \cite{Hongisto2006AAuwA} and Legault \cite{Legault2009JoSaV} contemporarily. However, none of them are appropriate for the case studied herein.

Composite multipanel structures with attachments or absorption fillings are emphasized more. However, their absorption fillings are complicated; in most cases, they are or can be considered as porous materials. Therefore, two widely used models for porous media can be used, i.e., the Biot theory \cite{Biot1956TJotASoA} and the equivalent fluid model (EFM) \cite{Allard1992TJotASoA}. In these absorption filling (cavity) problems, the EFM, owing to its simplicity, is widely used together with numerical \cite{Doutres2010TJotASoA} or semi-analytical methods \cite{Legault2009JoSaV,Trochidis1986JoSaV,Legault2010JoSaV}. 
For elastic frame porous problems, the Biot theory should be used \cite{Alimonti2015TJotASoA,Liu2017JoSaV} as the EFM is invalid.  Using the Biot theory, together with the simplifications of Deresiewicz \cite{Deresiewicz1962BotSSoA} and Allard \cite{Allard1989JoAP}, Bolton \cite{Bolton1996JoSaV} studied a two-dimensional (2D) multipanel structure with elastic porous materials, where the closed form expressions for 2D poroelastic field are obtained. The three-dimensional (3D) counterpart, with closed-form poroelastic field expressions, has been revealed by Zhou \cite{Zhou2013JoSaV}.  The effect of flow on these structures was subsequently studied by Liu \cite{Liu2015JoSaV}.
The numerical methods \cite{Alimonti2015TJotASoA} for these structures, based on the Biot theory, were also developed. 
 
Meanwhile, multipanel structures with attachments were prominent as well. The focus of the current ongoing study, as the absorption fillings are always absent under the circumstances, is primarily on those with ribs, resilient mountings, or elastic coatings. One of the most useful methods for these problems is the space harmonic series (SHS) introduced by Mead and Pujara \cite{Mead1971JoSaV}. It is widely used, when periodic ribs \cite{Xin2010JotMaPoS}, structure links \cite{Legault2009JoSaV} or resilient mountings \cite{Legault2010JoSaV} are present in the double-panel structure. The drawback of SHS was reported by Legault \cite{Legault2011JoSaV}. The Fourier transform method (FTM) \cite{Mace1980JoSaV} is also useful for these periodic problems. Multipanel structures, with absorption fillings \cite{Xin2010CSaT} (or not) \cite{Lin1977TJotASoA,Brunskog2005TJotASoA}, were studied using the FTM. In fact, as reported by Mace \cite{Mace1980JoSaV}, the same nature is shared between the FTM and SHS. Another useful method for these structures is the modal approach. Ribbed structures with simply supported condition, regarding their vibration \cite{Chung2008AAuwA}, structural intensity \cite{Brunskog2011TJotASoA}, and modal characteristics \cite{Dickow2013TJotASoA}, were studied using the modal approach with the appropriate orthonormal modal functions. Although applications in the double panel \cite{Liu2017JoSaV,Xin2009TJotASoA} can be found, however, relevant orthonormal modal functions for porous media are not available currently. 

Despite fruitful research reported on multipanel structures, studies regarding the combination of periodic structures with porous materials are scarce. Furthermore, in previous works, rather than the Biot theory, the EFM was used \cite{Legault2009JoSaV,Legault2010JoSaV,Xin2010CSaT} instead. In the present work, we study the sound insulation of a periodically rib-stiffened double panel with porous lining, to combine periodic structures and poroelastic materials. The periodic response are formulated in the SHS; meanwhile, based on the work of Bolton \cite{Bolton1996JoSaV} and Zhou \cite{Zhou2013JoSaV}, the periodic poroelastic field is obtained in the closed form using the Biot theory. A semi-analytical vibroacoustic model for the periodic composite structure can subsequently be established. It is solved by adopting the preconditioning method by Hull \cite{Hull2010JoSaV} with a truncation procedure. 
The novelty here is that the closed-form periodic poroelastic field, which is obtained for the first time, can be used to solve poroelastic problems with periodic boundary conditions semi-analytically. 

In Section \ref{stn:theory}, detailed model configurations are presented. The bonded--bonded case is described as an example and the solution procedures are outlined. Subsequently, the validation and parameter analyses are provided in Section \ref{stn:results}; Section \ref{stn:conclusion} ends with the conclusions.

\section{Modeling procedures for the periodic composite structure}\label{stn:theory}

\begin{figure}
     \centering
     \includegraphics[width=.8\textwidth]{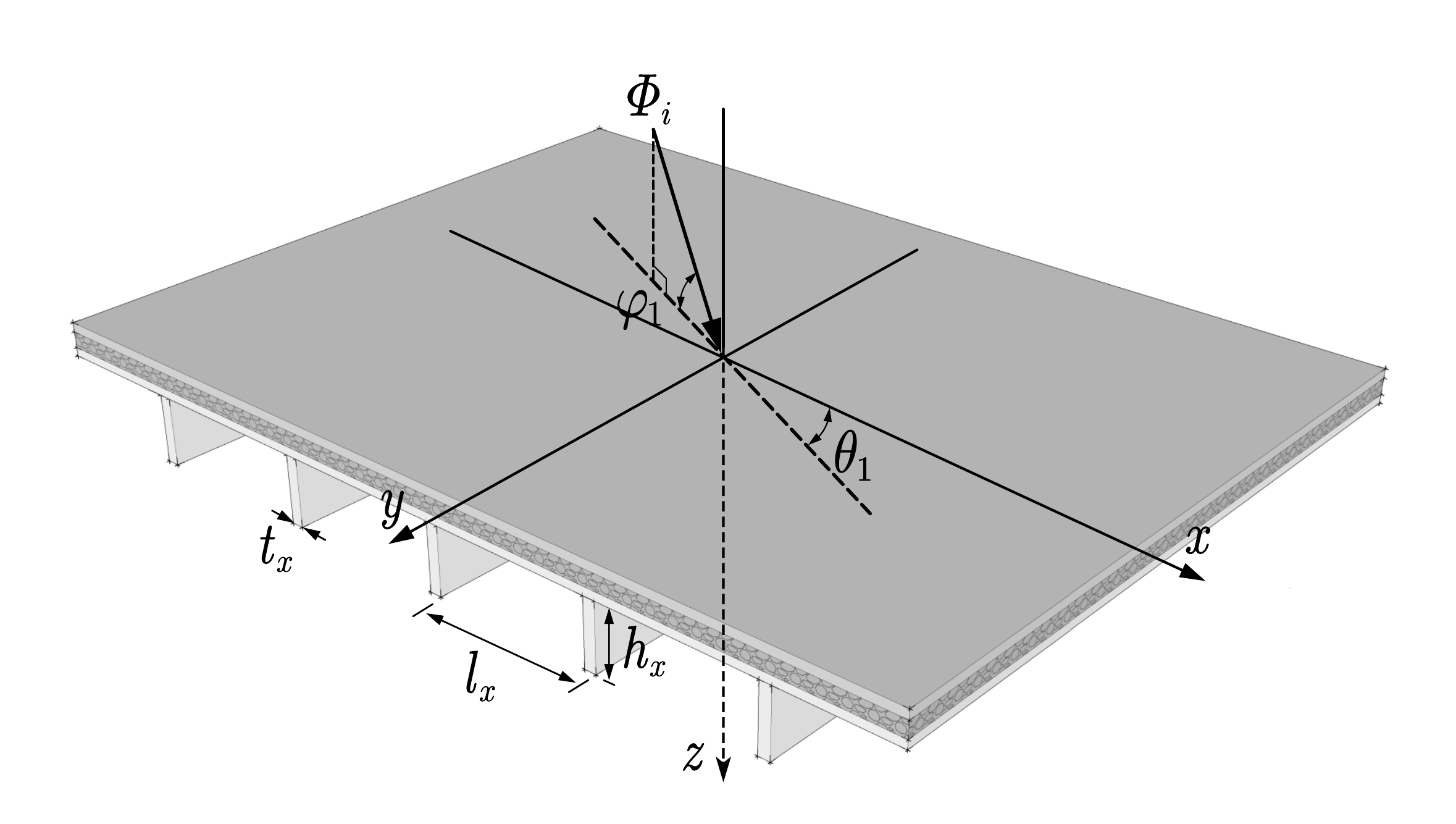}
     \caption{Schematic diagram of the periodic composite model (the ribs are exaggerated)}\label{fig:model}
 \end{figure}

The periodic composite structure is composed of a rib-stiffened double panel with porous lining, as shown in Fig.\ref{fig:model}; it is immersed in an inviscid stationary acoustic fluid. An incident wave $\varPhi_i= {\rm e}^{{\rm j}\omega t-{\rm j} \bf{k} \bf{r}}$ transmits through the structure, ${\bf k}=(k_x,k_y,k_z)$ is the incident wave vector,  ${\bf r}=(x,y,z)$, ${\rm j}=\sqrt{-1}$. According to Fig.\ref{fig:model}, $k_x=k_i {\rm cos}\varphi_1 {\rm cos}\theta_1,\ k_y=k_i {\rm cos}\varphi_1 {\rm sin}\theta_1,\ k_z=k_i {\rm sin}\varphi_1$; here, $k_i$ is the incident wavenumber, $\varphi_1$ and $\theta_1$ are the incident elevation angle and azimuth angle, respectively. The time-dependent term ${\rm e}^{{\rm j}\omega t}$ is omitted henceforth as the incident wave is time harmonic \cite{Legault2009JoSaV,Mace1981JoSaV}. The space occupied by the acoustic fluid on both sides is assumed to be semi-infinite and lossless; the density and sound velocity are designated as $\rho_i$,$c_i$ and $\rho_t$,$c_t$ for the incident and transmitted sides, respectively. 

The ribs are periodically placed along x at a spacing $l_x$, and extend infinitely along y; the thickness and height are $t_x$ and \ $h_x$, respectively (as shown in Fig.\ref{fig:model}). The longitudinal deformation of ribs is ignored as the y dimension is infinite.

\subsection{Velocity potentials in acoustic media}

As periodic cavities between the ribs are formed, SHS is used \cite{Xin2010JotMaPoS} to express the velocity potential $\varPhi_1$ of the incident side
\begin{equation}\label{eq:phi1}
\varPhi_1=\varPhi_i+\varPhi_r= {\rm e}^{-{\rm j} \bf{k} \bf{r}}+\sum_{m } R_m {\rm e}^{-{\rm j} {\bf k}_i^{m} {\bf r}}
\end{equation}
where ${\bf k}_i^{m}=(k_{i,x}^{m},k_{i,y}^{m},k_{i,z}^{m})$ is the wave vector; $R_m$ is the unknown amplitude of reflected wave harmonics; $k_{i,x}^{m}=k_x+{2 m\pi}/{l_x}$, 
$k_{i,y}^{m}=k_y$; integer $m\in [-\infty,\infty]$.
According to the wave equation ${\partial^2 \varPhi_1}/{\partial t^2}=c_i^2 \nabla^2 \varPhi_1$, $k_i={\omega}/{c_i}$, $k_{i,z}^{m}=\sqrt{k_i^2-\left(k_{i,x}^{m}\right)^2-(k_{i,y}^{m})^2}$.

The velocity potential of the transmitted side can be expressed as
\begin{equation}
\varPhi_2=\sum_{m} T_m {\rm e}^{-{\rm j} {\bf k}_{t}^{m} {\bf r}}\label{eq:phi2}
\end{equation}
Here, ${\bf k}_t^{m}=(k_{t,x}^{m},k_{t,y}^{m},k_{t,z}^{m})$ is the wave vector; $T_m$ is the unknown amplitude of the transmitted wave harmonics; integer $m\in[-\infty,\infty]$. According to the law of refraction \cite{Morse1968}, the wave vector component $k_{t,x}^{m}=k_{i,x}^{m}$,\ $k_{t,y}^{m}=k_{i,y}^{m}$. Substituting Eq.(\ref{eq:phi2}) into the wave equation, we obtain $k_{t,z}^{m}=\sqrt{k_t^2-\left(k_{t,x}^{m}\right)^2-\left(k_{t,y}^{m}\right)^2}$, $k_t=\omega/c_t$.

The corresponding sound pressure $p$ and acoustic particle velocity ${\bf v}$ can be obtained by $p=\rho\frac{\partial \varPhi}{\partial t}$, ${\bf v}=-\nabla\varPhi$, once the velocity potential $\varPhi$ is determined.

\subsection{In-plane and transverse vibration of face panels}
The two face panels are considered as isotropic thin plates. Their thicknesses, and displacements along x, y, and z are denoted as $h_i,u_i,v_i,w_i$, respectively; here, $i=1,2$ correspond to the incident and transmitted side panel. When the in-plane force and moment are present, the vibration equations are \cite{Axisa2005}
\begin{equation}
\mathcal{L}_i(u,v)=f_x ,\quad \mathcal{L}_t(w)=f_z -\frac{\partial \mathcal{M}_x }{\partial y}+\frac{\partial \mathcal{M}_y }{\partial x}
\end{equation}
where
\begin{gather}
\mathcal{L}_i(u,v)=\rho h  \frac{\partial^2 u }{\partial t^2}-D_p \frac{\partial}{\partial x}\left( \frac{\partial u }{\partial x}+\nu \frac{\partial v }{\partial y} \right)-G  h \frac{\partial}{\partial y}\left( \frac{\partial u}{\partial y}+ \frac{\partial v }{\partial x} \right)\\
\mathcal{L}_t(w)=D \nabla^4 w + \rho  h  \frac{\partial^2 w }{\partial t^2}
\end{gather}
are the in-plane vibration and transverse vibration operators, respectively; $f_z, f_x $ are the external forces along z and x, respectively; the in-plane moment ${\bf m} =\mathcal{M}_x {\bf i}_x+\mathcal{M}_y {\bf i}_y$, ${\bf i}_x$, and ${\bf i}_y$ are unit vectors along x and y, respectively; $\rho$ is the density, $h$ is the thickness, $D_p$ is the in-plane stiffness, $G$ is the shear modulus, and $D$ is the bending stiffness. The panel displacement ${\bf u}_i=(u_i,v_i,w_i)$ is expressed as
\begin{equation}
{\bf u}_i=\sum_{m}{\bf U}_i^m {\rm e}^{-{\rm j} ({k}_{x}^{m}x+{k}_{y}^{m}y)}\label{eq:panel_disp}
\end{equation}
Here, ${k}_{x}^{m}={k}_{i,x}^{m}$, ${k}_{y}^{m}={k}_{i,y}^{m}$ according to the law of refraction; ${\bf U}_i^m=(U_i^m,V_i^m,W_i^m)$ is the unknown component amplitude vector; integer $m\in [-\infty,\infty]$.
 
\subsection{Flexural vibration and rotation of ribs}
The flexural vibration of the ribs is modeled by the Bernoulli--Euler model and Timoshenko model in this study; a comparison between them is performed. The rotation of the ribs is modeled by the torsional wave equation.

The Bernoulli--Euler beam (BE--B) equation is
 \cite{Junger1986}
\begin{equation}
EI\frac{\partial^4 w}{\partial y^4}+\rho A \frac{\partial^2 w}{\partial t^2}=f_z \nonumber
\end{equation}
where $w$ is the displacement, $\rho$ is the density, $E$ is the Young's modulus, $I$ is the second moment of area, $A$ is the cross-section area, and $f_z$ is the external force. 
The Timoshenko beam (TS--B) equation is
 \cite{Junger1986}
\begin{equation}
EI\frac{\partial^4 w}{\partial y^4}- \rho I\left(1+\frac{ E}{G \kappa} \right)\frac{\partial^4 w}{\partial y^2 \partial t^2} +\rho A\frac{\partial^2 w}{\partial t^2}+\frac{\rho^2 I}{G\kappa}\frac{\partial^4 w}{\partial t^4} =f_z-\frac{E I}{G A \kappa}\frac{\partial^2 f_z}{\partial y^2}+\frac{\rho I}{G A \kappa}\frac{\partial^2 f_z}{\partial t^2}\nonumber 
\end{equation}
where $G$ is the shear modulus, $\kappa$ is the shear correction factor. 

The rotation is determined by
 \cite{Rao2007}
\begin{equation}
G I_p\frac{\partial^2 \theta_y}{\partial y^2}-\rho I_p\frac{\partial^2 \theta_y}{\partial t^2}=m_y \nonumber 
\end{equation}
where $I_p$ is the polar moment of inertia, $m_y$ is the external moment; $\theta_y={\partial w_2}/{\partial x}$ is the clockwise angle of rotation about the rib-panel interface. 
Here, $w_2$ is the panel displacement.

Subsequently, the forces exerted on the transmitted side plate can be obtained using the equations above and the displacement continuity condition. The results are 
\begin{equation}
f_z=K_z w_2,\ m_y=K_y \sum_{m} {\rm j} {k}_{x}^{m} W_2^m {\rm e}^{-{\rm j} ({k}_{x}^{m}x+{k}_{y}^{m}y)}\label{eq:rib_single_force}
\end{equation}
here $K_y=G I_p k_y^2-\rho I_p \omega^2$, while 
\begin{equation}
K_z=\begin{cases}
E I k_y^4-\rho A \omega^2,& \text{For BE-B case}\\
\frac{E I k_y^4-\rho I(1+  E/G\kappa)k_y^2\omega^2-\rho A\omega^2+(\rho^2 I/G\kappa)\omega^4}{1+(EI/GA\kappa) k_y^2-(\rho I /G A\kappa)\omega^2},& \text{For TS-B case}
\end{cases}
\end{equation}
depending on the beam model.

The resultant force exerted by the periodic ribs can subsequently be written as
\begin{gather}
F_z=\sum_{n}f_z\delta(x-n l_x)+ \sum_{n}\frac{\partial m_y}{\partial x}\delta(x-n l_x)
\end{gather}
Here, $n\in [-\infty,\infty]$. It then becomes
\begin{equation}\label{eq:ribforces}
 F_z = \frac{1}{l_x}\sum_{n} \sum_{m} \left[K_z W_2^m +({k}_{x}^{m})^2 K_y W_2^m \right]{\rm e}^{-{\rm j}\big(k_x^{m} x+k_y^{m} y\big)} {\rm e}^{{\rm j}2\pi n x/l_x}
\end{equation}
utilizing Eq.(\ref{eq:rib_single_force}) and the Poisson summation formula \cite{Mace1981JoSaV}
\begin{equation}
\sum_{n=-\infty,\infty}\delta(x-n l_x)=\frac{1}{l_x}\sum_{n=-\infty,\infty}{\rm e}^{{\rm j} 2\pi n x/l_x}
\end{equation}

A double summation is present, which renders the problem complicated and cumbersome. An index separation identity is utilized to eliminate it subsequently. 

\subsection{Poroelastic field with periodic boundary conditions}

In poroelastic problems, the appropriate porous modeling technique is not available when periodic boundary conditions are present.
In this study, the field variables are assumed to consist of six groups of harmonics according to their wavenumbers, while only six components were considered formerly \cite{Bolton1996JoSaV,Zhou2013JoSaV}. The poroelastic field is subsequently obtained in terms of the harmonic coefficients, which can be solved with the proper boundary conditions. The porous material in this study is assumed to be isotropic with homogeneous cylindrical pores to obtain an elegant formulation.

The poroelastic equations expressed by solid and fluid displacements ${\bf u}^s$, ${\bf u}^f$, respectively, are \cite{Allard2009,Bolton1996JoSaV}
\begin{gather}
    -\omega^2\left({\rho}_{11}^{*}{\bf u}^{s}+{\rho}_{12}^{*} {\bf u}^{f}\right)=(A+N)\nabla\nabla\cdot{\bf u}^{s}+N\nabla^2{\bf u}^{s}+Q\nabla\nabla\cdot{\bf u}^{f}\label{eq:biot1}\\
    -\omega^2\left({\rho}_{22}^{*}{\bf u}^{f}+{\rho}_{12}^{*}{\bf u}^{s}\right)=R\nabla\nabla\cdot{\bf u}^{f} +Q\nabla\nabla\cdot{\bf u}^{s}\label{eq:biot2}
\end{gather}
where ${\rho}_{11}^{*}={\rho}_{11}+b/{\rm j}\omega$,\ ${\rho}_{12}^{*}={\rho}_{12}-b/{\rm j}\omega$,\ ${\rho}_{22}^{*}={\rho}_{22}+b/{\rm j}\omega$,\ ${\rho}_{11} ={\rho}_{1}+{\rho}_{a}$,\ ${\rho}_{12}=-{\rho}_{a} ={\rho}_{2}(1-\epsilon')$,\ ${\rho}_{22} ={\rho}_{2}+{\rho}_{a}$, $b={\rm j}\omega\epsilon'\rho_2({\rho}_{c}^{*}/\rho_f-1)$,\ ${\rho}_{c}^{*}=\rho_f [1-2 T_c(\lambda_1)/\lambda_1 ]^{-1}$,\ $\lambda_1=\lambda_c\sqrt{-{\rm j}}$,\ $\lambda_c=\sqrt{8\omega\rho_f\epsilon'/\phi\sigma}$; $\rho_f$ is the density of ambient fluid; $\phi$ is the porosity; $\rho_1=\rho_s$ and $\rho_2=\phi\rho_f$ are the bulk solid and fluid densities; $\epsilon'$ is the tortuosity; $\sigma$ is the flow resistivity; auxiliary function $T_c(x)=J_1(x)/J_0(x)$, $J_1(x)$, and $J_0(x)$ are Bessel functions of the first kind, first and zero order, respectively. Parameter $A=\nu_s E_s/(1+\nu_s)(1-2\nu_s)$ is the first Lame constant; $N=E_s/2(1+\nu_s)$ is the shear modulus; the coupling parameter $Q=(1-\phi)E_f$, $R=\phi E_f$; here, $E_f=\rho_f c_f^2[1+2(\gamma-1)T_c(\lambda_2)/\lambda_2]^{-1}$ is the bulk modulus of the fluid in the pores, $c_f$ is the sound velocity of the fluid in pores, $\gamma$ is the ratio of specific heats and auxiliary variable $\lambda_2=\lambda_c\sqrt{-{\rm j} N_{\rm Pr}}$, and $N_{\rm Pr}$ is the Prandtl number in the pores.

The poroelastic equations of Eqs.(\ref{eq:biot1})-(\ref{eq:biot2}) can be reduced to two wave equations \cite{Bolton1996JoSaV,Zhou2013JoSaV} (a fourth-order equation and a second-order equation) when two scalar potentials $\varphi^s=\nabla\cdot {\bf u}^s,\ \varphi^f=\nabla\cdot{\bf u}^f$ and two vector potentials ${\bf \Psi}^s=\nabla\times{\bf u}^s,\ {\bf \Psi}^f=\nabla\times{\bf u}^f$ are introduced; these potentials are the dilatational and rotational strains of the corresponding phases \cite{Allard2009,Bolton1996JoSaV}. The wavenumbers corresponding to the wave equations are
\begin{equation}
\{k_{1}^2,k_{2}^2\}=A_1/2\pm\sqrt{A_1^2/4-A_2},\ k_3^2=\omega^2/N({\rho}_{11}^{*}-{\rho}_{12}^{*}{\rho}_{12}^{*}/{\rho}_{22}^{*})
\end{equation}
Here, the auxiliary term $A_1=\omega^2({\rho}_{11}^{*}R-2{\rho}_{12}^{*}Q+{\rho}_{22}^{*}P)/(PR-Q^2)$, $A_2=\omega^4({\rho}_{11}^{*}{\rho}_{22}^{*}-{\rho}_{12}^{*}{\rho}_{12}^{*})/(PR-Q^2)$, $P=A+2N$.

Utilizing SHS, the solid phase strain $e^s=\varphi^s$, ${\bf \Omega}^s={\bf \Psi}^s$ are written as
\begin{gather}
    e^s=\sum_{m}{\rm e}^{-{\rm j}\left(k_x^m x+k_y^m y\right)} \left(C_1^{m} {\rm e}^{-{\rm j} k_{1,z}^{m} z}+C_2^{m} {\rm e}^{{\rm j} k_{1,z}^{m} z} +C_3^{m} {\rm e}^{-{\rm j} k_{2,z}^{m} z}+C_4^{m} {\rm e}^{{\rm j} k_{2,z}^{m} z} \right)\label{eq:strains1}
 \\
    |{\bf\Omega}^s|=\sum_{m}{\rm e}^{-{\rm j}\left(k_x^{m} x+k_y^{m} y\right)}\left(C_5^{m} {\rm e}^{-{\rm j} k_{3,z}^{m} z}+C_6^{m} {\rm e}^{{\rm j} k_{3,z}^{m} z} \right)\label{eq:strains2}
\end{gather}
Here, $C_i^m\ (i=1,2...6)$ are the unknown amplitude of the harmonic components; $k_{1,z}^m$, $k_{2,z}^m$, $k_{3,z}^m$ are the z component of the corresponding wave vectors; the law of refraction is used here. According to Eqs.(\ref{eq:biot1})-(\ref{eq:biot2}), the fluid phase strain $e^f=\varphi^f$, ${\bf\Omega}^f={\bf \Psi}^f$ are
\begin{gather}
    e^f=\sum_{m}{\rm e}^{-{\rm j}\left(k_x^m x+k_y^m y\right)} \left(b_1^{m} {\rm e}^{-{\rm j} k_{1,z}^{m} z}+b_2^{m} {\rm e}^{{\rm j} k_{1,z}^{m} z} +b_3^{m} {\rm e}^{-{\rm j} k_{2,z}^{m} z}+b_4^{m} {\rm e}^{{\rm j} k_{2,z}^{m} z} \right)\label{eq:strainf1}
 \\
    |{\bf\Omega}^f|= g {\bf\epsilon}^s \label{eq:strainf2}
\end{gather}
where $b_1^m=(a_1-a_2 k_1^2)C_1^{m}$, $b_2^m=(a_1-a_2 k_1^2)C_2^{m}$, $b_3^m=(a_1-a_2 k_2^2)C_3^{m}$, $b_4^m=(a_1-a_2 k_2^2)C_4^{m}$, $g=-{\rho}_{12}^{*}/{\rho}_{22}^{*}$; the auxiliary term $a_1=({\rho}_{11}^{*}R-{\rho}_{12}^{*}Q)/({\rho}_{22}^{*}Q-{\rho}_{12}^{*}R)$, $a_2=(P R-Q^2)/\omega^2({\rho}_{22}^{*}Q-{\rho}_{12}^{*}R)$.
Using the definition of potentials and substituting Eqs.(\ref{eq:strains1})-(\ref{eq:strainf2}) into Eqs.(\ref{eq:biot1})-(\ref{eq:biot2}), subsequently $k_{i,z}^{m}=\sqrt{k_i^2-(k_x^{m})^2-(k_y^{m})^2}$ ($i$=1,2,3) can be obtained.

The field variables are composed of six groups of harmonics (i.e., $\pm k_{1,z}^m$, $\pm k_{2,z}^m$, $\pm k_{3,z}^m$) here.
According to Bolton \cite{Bolton1996JoSaV} with $\nabla\cdot{\bf\Omega}^s=0$ and $\nabla\cdot{\bf\Omega}^f=0$ \cite{Graff1975}, the poroelastic displacement ${\bf u}=[{u}_x^s,{u}_y^s,{u}_z^s,{u}_x^f,{u}_y^f,{u}_z^f]^T$ is obtained
\begin{gather}\label{eq:porousvelocity}
{\bf u}=\sum_{m}{\rm e}^{-{\rm j}\big(k_x^{m} x+k_y^{m} y\big)}\ {\bf Y}_m{\bf e}_m {\bf C}_m
\end{gather}
where 
\begin{gather}
{\bf e}_m =diag({\rm e}^{-{\rm j} k_{1,z}^{m} z},{\rm e}^{{\rm j} k_{1,z}^{m} z},{\rm e}^{-{\rm j} k_{2,z}^{m} z},{\rm e}^{{\rm j} k_{2,z}^{m} z},{\rm e}^{-{\rm j} k_{3,z}^{m} z},{\rm e}^{{\rm j} k_{3,z}^{m} z})\\
{\bf C}_m =[C_1^{m},C_2^{m},C_3^{m},C_4^{m},C_5^{m},C_6^{m}]^T
\end{gather}
the matrix ${\bf e}_m$ is a $6\times6$ diagonal matrix. The elements of the coefficient matrix ${\bf Y}_m$ are given in \ref{app:porousexpression}. The forces in the porous media are \cite{Biot1956TJotASoA,Allard2009}
\begin{gather}
    \sigma_{ij}=2N e_{ij}+(A e^s + Q e^f)\delta_{ij}\label{eq:porous_force1}\\
    s=Q e^s+R e^f\label{eq:porous_force2}
\end{gather}
where $\sigma_{ij}$ and $s$ are the forces in the solid and fluid phase, $\delta_{ij}$ is the Kronecker delta, $e_{ij}$ is the normal $(i=j)$ or shear ($i\ne j$) strain
\begin{equation}
    e_{ij}=\begin{cases}
    \partial u_i/\partial x_i,&  i = j \\
    \frac{1}{2}\left(\partial u_i/\partial x_j+\partial u_j/\partial x_i\right),&   i \ne j
    \end{cases},\quad
    \delta_{ij}=\begin{cases}
    1,&  i = j \\
    0,&  i \ne j
    \end{cases}
\end{equation}

\subsection{Boundary conditions}
The boundary condition notation of Bolton \cite{Bolton1996JoSaV} is used; the connection type can be bonded--bonded (BB), bonded--unbonded (BU), and unbonded--unbonded (UU). The related boundary conditions are not presented herein, as they can be found in Bolton \cite{Bolton1996JoSaV}, Zhou \cite{Zhou2013JoSaV} and Liu \cite{Liu2015JoSaV}, etc.
It is noteworthy that a right-handed coordinate system is used by Liu and herein, while Bolton and Zhou used the left-handed one. The detailed expressions are identical for the 2D case; however, in the 3D case, some modifications should be performed between the two coordinate systems.

\subsection{System equations and solution procedures}
The BB case is used as an example to demonstrate the solution procedure. The related boundary conditions are 
\begin{gather}
({\rm i})\ -{\rm j}\omega\frac{\partial \varPhi_1}{\partial z}=\frac{{\rm \partial}^2 w_1}{{\rm \partial} t^2}\quad ({\rm ii})\ \mathcal{L}_i(u_1,v_1)=\tau_{zx}\ \notag\\
({\rm iii})\ \mathcal{L}_t(w_1)={\rm j}\omega\rho_i\varPhi_1+(\sigma_z+s)-\frac{h_1}{2}\left(\frac{\partial\tau_{zx}}{\partial x}+\frac{\partial\tau_{zy}}{\partial y} \right) \notag\\
({\rm iv})\ u_z^s=w_1\quad ({\rm v})\ u_z^f=w_1\quad ({\rm vi})\ u_x^s=u_1-\frac{h_1}{2}\frac{\partial w_1}{\partial x}\notag\\
({\rm vii})\ u_y^s=v_1-\frac{h_1}{2}\frac{\partial w_1}{\partial y}\quad ({\rm viii})\ u_z^s=w_2\quad ({\rm ix})\ u_z^f=w_2\notag\\
({\rm x})\ u_x^s=u_2+\frac{h_2}{2}\frac{\partial w_2}{\partial x}\quad ({\rm xi})\ u_y^s=v_2+\frac{h_2}{2}\frac{\partial w_2}{\partial y}\notag\\
({\rm xii})\ \mathcal{L}_i(u_2,v_2)=-\tau_{zx}\notag\\
({\rm xiii})\ \mathcal{L}_t(w_2)=-{\rm j}\omega\rho_t\varPhi_2-(\sigma_z+s)-\frac{h_2}{2}\left(\frac{\partial\tau_{zx}}{\partial x}+\frac{\partial\tau_{zy}}{\partial y} \right)-F_z\notag\\
({\rm xiv})\ -{\rm j}\omega\frac{\partial \varPhi_2}{\partial z}=\frac{\partial^2 w_2}{\partial t^2}\label{eq:sysbb}
\end{gather}
where $F_z$ is the resultant force exerted by the ribs in Eq.(\ref{eq:ribforces}); (i)--(xiv) are applied on the domain interfaces or panel middle surfaces \cite{Bolton1996JoSaV,Zhou2013JoSaV}.

To eliminate the summation index used, the orthogonal property below
\begin{equation}\label{eq:orthomp}
\int_{-l_x/2}^{l_x/2} {\rm e}^{-{\rm j} ({k}_{x}^{m}x+{k}_{y}^{m}y)} {\rm e}^{{\rm j} ({k}_{x}^{p}x+{k}_{y}^{p}y)} dx = 
\begin{cases} 
l_x,\  & m=p \\
0,\ & m\ne p
\end{cases}
\end{equation}
and a double summation identity
\begin{equation}\label{eq:identity}
\sum_{n=-\infty}^{+\infty}\sum_{m=-\infty}^{+\infty}W_m {\rm e}^{-{\rm j} {k}_{x}^{m}x} {\rm e}^{{\rm j} 2\pi n x/l_x}=\sum_{n=-\infty}^{+\infty}W_n \sum_{m=-\infty}^{+\infty} {\rm e}^{-{\rm j} {k}_{x}^{m}x}
\end{equation}
are used. The derivation of Eq.(\ref{eq:identity}) is given in \ref{app:identityproof}.

Substitute Eqs.(\ref{eq:phi1}),\ (\ref{eq:phi2}),\ (\ref{eq:panel_disp}),\ (\ref{eq:ribforces}),\ (\ref{eq:porousvelocity}), (\ref{eq:porous_force1}) and (\ref{eq:porous_force2}) into (\ref{eq:sysbb}), and utilize Eqs.(\ref{eq:orthomp})-(\ref{eq:identity}), Eq.(\ref{eq:sysbb}) becomes
\begin{equation}\label{eq:sysmatrix}
{\bf A}_m {\bf x}_m=\sum_{n} {\bf B}_n {\bf x}_n +
\begin{cases} 
{\bf p},\  & m=0 \\
{\bf 0},\ & m\ne 0
\end{cases}
\end{equation}
where
\begin{gather}
{\bf x}_m=[C_1^{m},C_2^{m},C_3^{m},C_4^{m},C_5^{m},C_6^{m},U_1^m,V_1^m,W_1^m,U_2^m,V_2^m,W_2^m,R_m,T_m]^T \notag
\end{gather}
Here, ${\bf 0}$ is a zero $14\times 1$ vector; all the elements of the corresponding matrices and vectors in Eq.(\ref{eq:sysmatrix}) are given in \ref{app:matrixelements}. Equation (\ref{eq:sysmatrix}) should be solved for every integer $m, n\in[-\infty, +\infty]$, and is thus a matrix system of infinite dimension.

The Eq.(\ref{eq:sysmatrix}) is rearranged using a procedure analogous to Hull \cite{Hull2010JoSaV} to obtain a solution. 
The first step is to rearrange ${\bf x}_m$ into 
\begin{equation}
\tilde{\bf x}=[\cdots,\ {\bf x}_{m-1}^T,\ {\bf x}_{m}^T,\ {\bf x}_{m+1}^T,\ \cdots]^T
\end{equation}
Subsequently, the matrix ${\bf A}_m$ can be rearranged to a block diagonal matrix
\begin{equation}
\tilde{\bf A}=
\begin{bmatrix}
\ddots &   &  & &      \\
 & {\bf A}_{m-1} & & & \\
 &  &{\bf A}_m& & \\
 &  & &{\bf A}_{m+1} & \\
  &  & &  & \ddots 
\end{bmatrix}
\end{equation}
Here, the blank elements of $\tilde{\bf A}$ are all zero; the matrix ${\bf B}_n$ can be rearranged to a full block matrix
\begin{equation}\label{eq:matrix_bmd}
\tilde{\bf B}=
\begin{bmatrix}
\ddots &   &  & & \\
\cdots & {\bf B}_{n-1} &{\bf B}_{n} &{\bf B}_{n+1} &\cdots  \\
\cdots & {\bf B}_{n-1} &{\bf B}_{n} &{\bf B}_{n+1} &\cdots \\
\cdots & {\bf B}_{n-1} &{\bf B}_{n} &{\bf B}_{n+1} &\cdots \\
 &  & &  & \ddots 
\end{bmatrix}
\end{equation}
where the blank elements of $\tilde{\bf B}$ are in the same pattern as shown in Eq.(\ref{eq:matrix_bmd}); the vector ${\bf p}$ can be rearranged to 
\begin{equation}
\tilde{\bf p}=[\cdots,\ {\bf 0}^T,\ {\bf p}^T,\ {\bf 0}^T,\ \cdots]^T
\end{equation}
Here, ${\bf 0}^T$ is a zero $1\times 14$ vector. 
Subsequently, Eq.(\ref{eq:sysmatrix}) can be written as 
\begin{equation}\label{eq:finalmatrix}
\tilde{\bf A}\tilde{\bf x}=\tilde{\bf B}\tilde{\bf x} + \tilde{\bf p}
\end{equation}

Equation (\ref{eq:finalmatrix}) needs to be truncated to obtain a solution, i.e., truncate the index number in $\tilde{\bf A}$, $\tilde{\bf B}$ and $\tilde{\bf x}$ to $[-\hat{m},\hat{m}]$; considerable accuracy can be ensured by the appropriate convergence criteria. Subsequently, the matrices $\tilde{\bf A}$, $\tilde{\bf B}$ are reduced to $14\hat{M} \times 14\hat{M}$ , and the vectors $\tilde{\bf x}$, $\tilde{\bf p}$ are reduced to $14\hat{M} \times 1$, $\hat{M}=2\hat{m}+1$; thus, Eq.(\ref{eq:finalmatrix}) can be solved using $\tilde{\bf x}=(\tilde{\bf A}-\tilde{\bf B})^{-1} \tilde{\bf p}$.

The sound field here can be regarded as the sum of all harmonics.
Thus, the transmission coefficient $\tau$ is \cite{Legault2009JoSaV}
\begin{equation}\label{eq:tauphi1theta1}
\tau(\varphi_1,\theta_1)=\frac{\rho_t}{\rho_i}\frac{\sum_{m} \left| T_m \right|^2 {\rm Re}(k_{t,z}^{m})}{{\rm Re}(k_z)}
\end{equation}
Here, ${\rm Re}(\cdot)$ is the real operator of a complex variable.

The random STL can subsequently be expressed as 
\cite{Zhou2013JoSaV}
\begin{gather}
{\rm STL}=10\ {\rm log}(1/\bar{\tau}),\quad \bar{\tau}=\frac{\int_{0}^{2\pi}\int_{\varphi_{\rm min}}^{\pi/2}\tau(\varphi_1,\theta_1) {\rm sin}\varphi_1{\rm cos}\varphi_1\ {\rm d}\varphi_1{\rm d}\theta_1}{\int_{0}^{2\pi}\int_{\varphi_{\rm min}}^{\pi/2}{\rm sin}\varphi_1{\rm cos}\varphi_1{\rm d}\varphi_1{\rm d}\theta_1}\label{eq:randomstl}
\end{gather}
Here, $\varphi_{\rm min}$ is the minimum elevation angle.

\section{Results and discussions}\label{stn:results}

This part begins with a discussion of the convergence characteristics; subsequently, the validation is conducted. Several parameter analyses are performed finally. The rectangular ribs (aluminum) and porous parameters given in \cite{Bolton1996JoSaV,Zhou2013JoSaV,Liu2015JoSaV} are used. The detailed values are listed in Table \ref{tab:parameters}; they are used if no other values are specified hereinafter. The random STL is calculated in 1/24 octave bands using the 2D Simpson rule; $\varphi_{\rm min}=\pi/10$ \cite{Bolton1996JoSaV} as no convective flow is present; the integration domain of $\varphi_1$ and $\theta_1$ are split into 36 and 90 subdivisions, respectively.

\begin{table}[!ht]
\caption{Parameters in the periodic composite structure: air gap thickness $h_{a}$ = 14 mm for the BU case; $h_{a1}$ = 2mm, $h_{a2}$ = 6 mm for the incident and transmitted sides in the UU case, respectively; the characteristic thicknesses are $d=h_p$, $(h_p+h_a)$, and $(h_p + h_{a1}+ h_{a2})$ for the BB, BU, and UU cases, respectively; the gap properties $\rho_g=\rho_i, c_g=c_i$ and the transmitted side media properties $\rho_t=\rho_i, c_t=c_i$}\label{tab:parameters}
\begin{tabular}{lll}
\hline
Parameters & Physical description& Value\\
\hline
Acoustic media\\
$\rho_i$ & density (incident side) & 1.205 kg/${\rm m}^{3}$\\
$c_i$ & sound velocity (incident side) & 343 m/${\rm s} $\\
\hline
Double-panels\\
$\rho_p$ & density of face panels & 2700 kg/${\rm m}^{3}$\\
$E_p$ & Young's modulus of face panels & 70$\times 10^9$Pa\\
$\nu_p$ & Poisson's ratio of face panels & 0.33\\
$h_1$ & panel thickness (incident side)  & 1.27 mm\\
$h_2$ & panel thickness (transmitted side) & 0.762 mm\\
$h_p$ & thickness of porous core & 27 mm\\
\hline
Ribs\\
$l_x$ & rib spacing along x & 50$d$ \\
$t_x$ & rib thickness & 1 mm\\
$h_x$ & rib height & 20 mm\\
\hline
Porous media\\
$\rho_s$ & bulk density of solid phase & 30 kg/${\rm m}^{3}$\\
$\rho_f$ & density of fluid phase & 1.205 kg/${\rm m}^{3}$\\
$E_s$ & Young's modulus (solid phase) & 8$\times 10^5$Pa\\
$\nu_s$ & Poisson's ratio (solid phase) & 0.4\\
$\eta_s$ & loss factor (solid phase) & 0.265\\
$\phi$ & the porosity & 0.9\\
$\epsilon'$ & the tortuosity & 7.8\\
$\sigma$ & flow resistivity & 2.5$\times 10^4$ MKS\ Rayls/${\rm m} $\\

\hline
\end{tabular}
\end{table}

\subsection{Discussion of the convergence characteristics}

\begin{figure}[!htb]
     \centering
     \includegraphics[width=.75\textwidth]{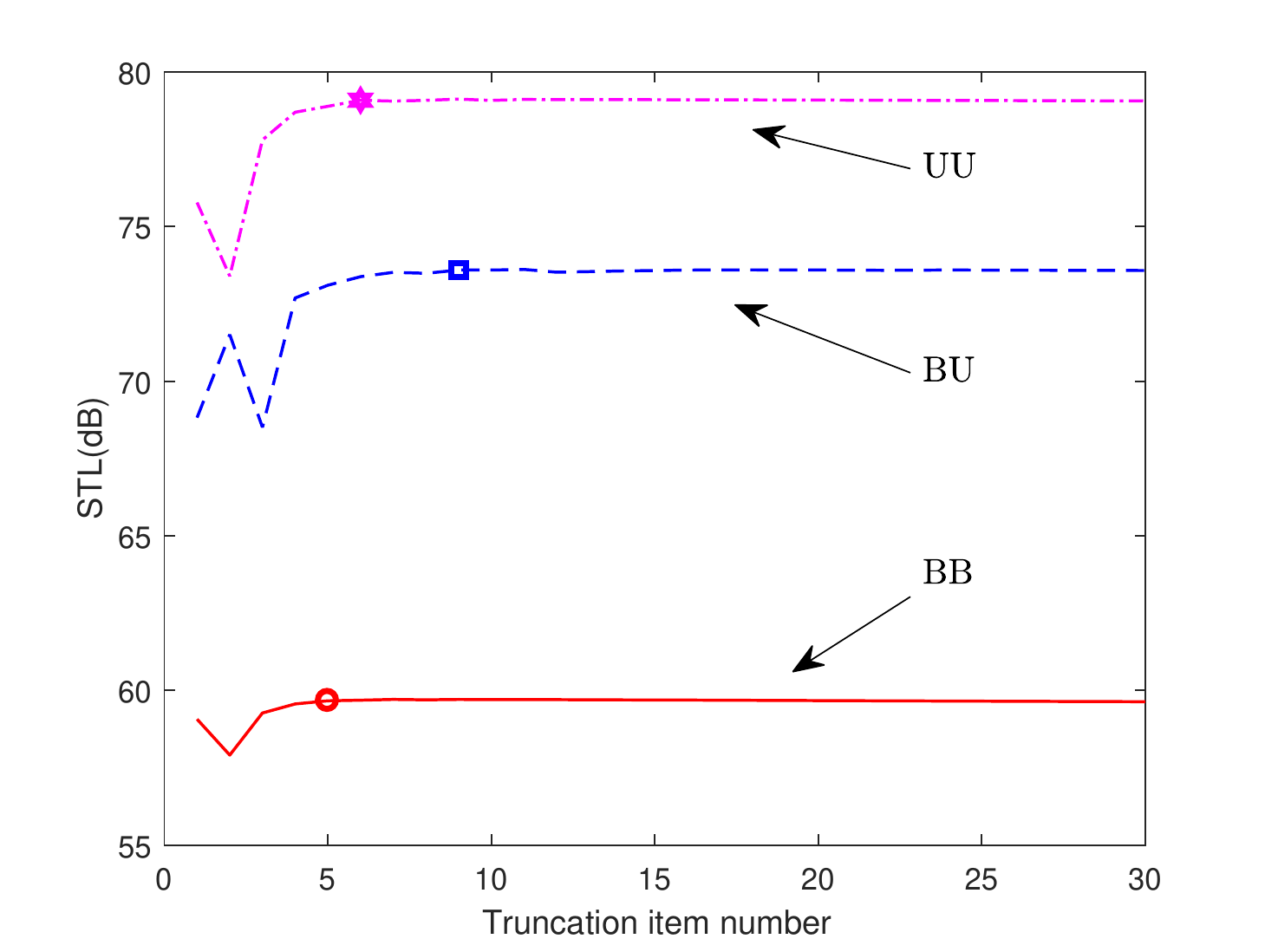}
     \caption{Variation in STL with truncation item number $\hat{m}$ at $f$=10kHz ($l_x$=50$d$)}\label{fig:m3dim}
\end{figure}

A truncation procedure is required to solve the infinite matrix equation Eq.(\ref{eq:finalmatrix}).
We set the convergence criteria as $\Delta {\rm STL}=$ 0.1 dB under the maximum computation frequency ($f$=10kHz); that is, when the change in STL by one additional $\hat{m}$ item is less than $\Delta {\rm STL}$, it is considered as converged. A typical convergence curve of the three boundary conditions is given in Fig.\ref{fig:m3dim}. Different marks (the circle, square, and hexagram for the BB, BU, and UU cases, respectively) herein show the convergence points in the given figure.

\begin{figure}[!htb]
     \centering
     \includegraphics[width=.8\textwidth]{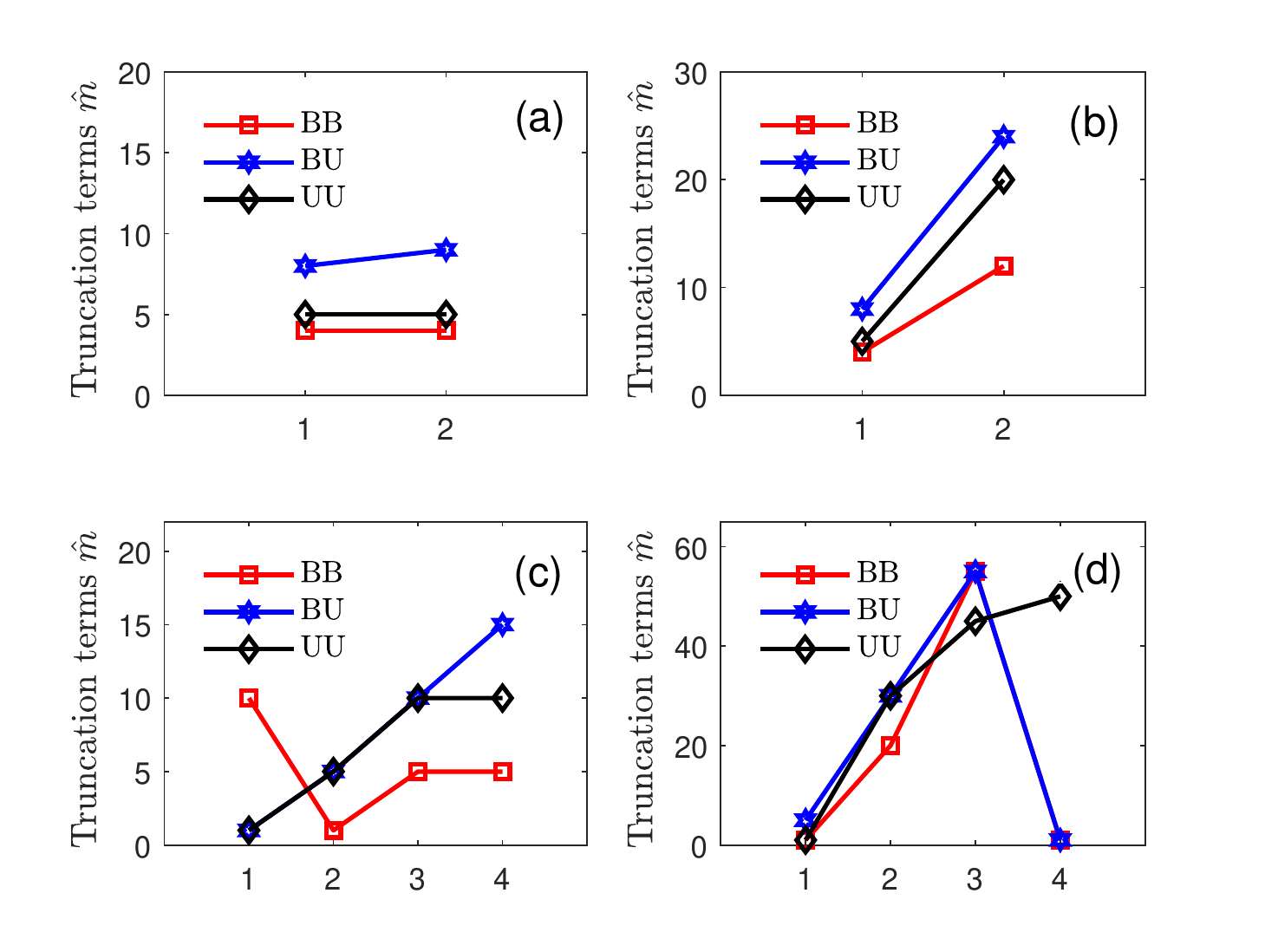}
     \caption{Truncation item number $\hat{m}$ required for (a) the rib model case (1 BE-B,\ 2 TS-B); (b) the torsion motion case (1-with,\ 2-w/o); (c) the rib spacing $l_x$ case (1--4 correspond to $l_x/d$=20, 50, 100, 150 respectively) ; (d) the area moment of inertia $I$ case (1-4 correspond to $t_x/h_x$=1/3, 1/5, 1/10, 1/20 respectively)}\label{fig:converge_im}
\end{figure}

The truncation item number $\hat{m}$ required for different parameter cases are shown in Fig.\ref{fig:converge_im}. 
As shown, $\hat{m}$ is dependent on both the boundary conditions and parameter values. A related study concerning $\hat{m}$ is in progress; however, no conclusion can be drawn currently. Therefore, one needs to determine $\hat{m}$ in every numerical case when the proposed method is used. This is the primary drawback of the proposed method.
 
\subsection{Model Validation}\label{sec:validation}
The poroelastic field expressions are validated using the theoretical results of the unribbed double-panel structure with porous lining (corresponds to $\tilde{\bf B}$ as a zero matrix and $l_x$ tends to infinity) by Bolton \cite{Bolton1996JoSaV} (2D), Zhou \cite{Zhou2013JoSaV} (3D) and Liu \cite{Liu2015JoSaV} (3D). The rib-stiffened configurations are validated with the FEM results, as no theoretical or experimental result can be found in the literature thus far, to the authors' knowledge.

\subsubsection{Validation of poroelastic field expressions}

\begin{figure}[!htb]%
     \centering
     \includegraphics[width=.8\textwidth]{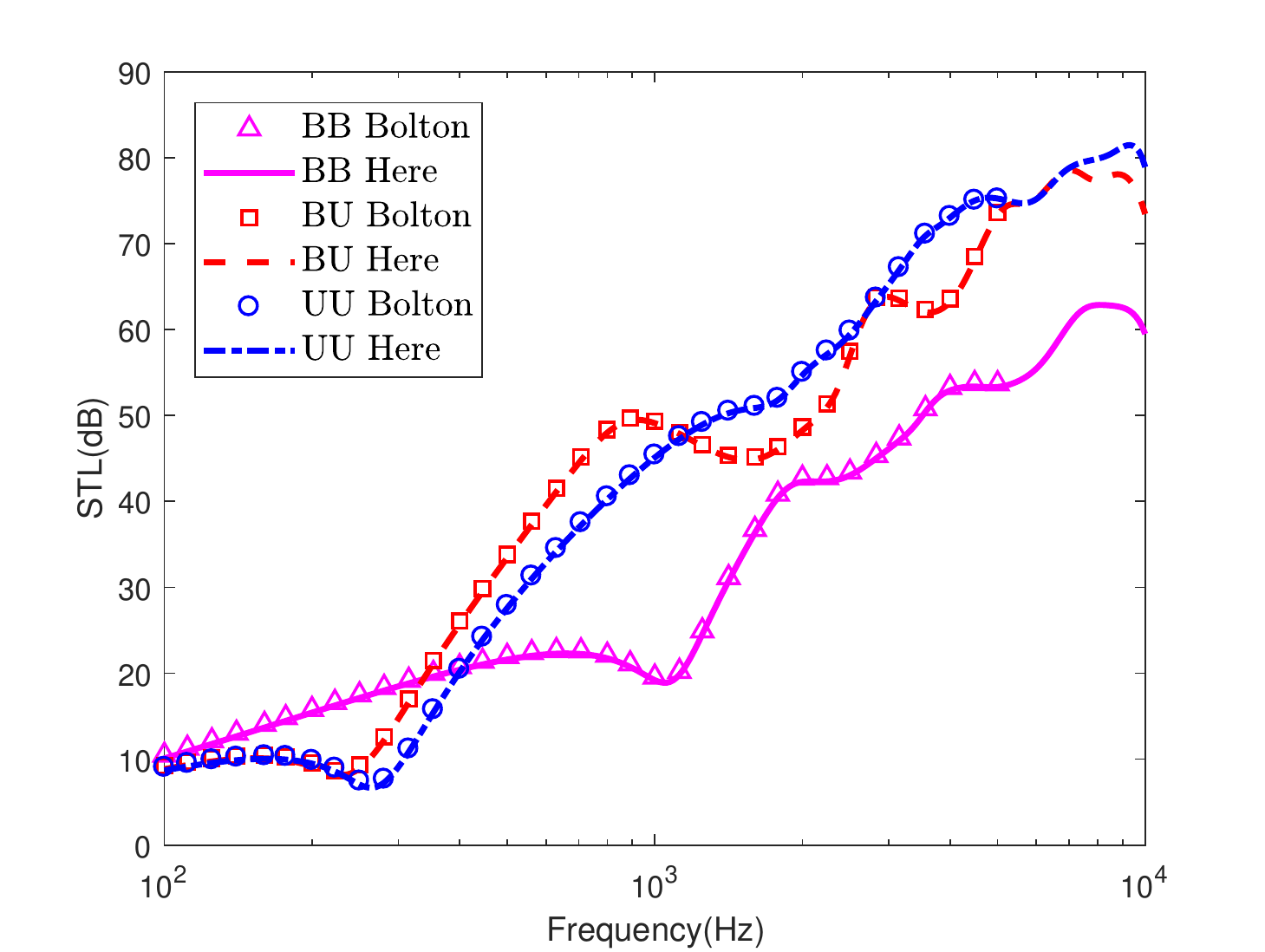}
     \caption{Validation of porous modeling with Bolton (2D)}\label{fig:val1bolton}
\end{figure}
 
\begin{figure}[!htb]
     \centering
     \includegraphics[width=.8\textwidth]{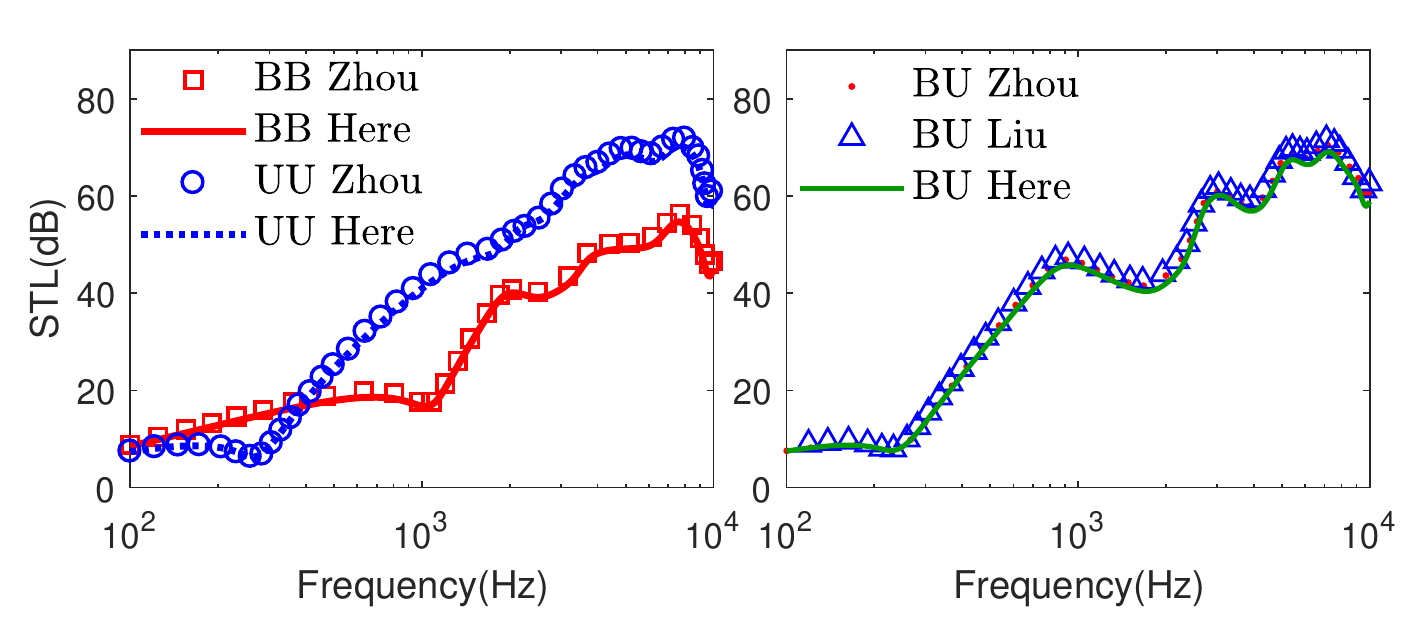}
     \caption{Validation of porous modeling with Zhou (3D) and Liu (3D)}\label{fig:valid3zhou}
\end{figure}

The model is first reduced to 2D by setting $\theta_1$=0. The comparison of results with Bolton are presented in Fig.\ref{fig:val1bolton} and the consistency is good. Subsequently, the validation with Zhou \cite{Zhou2013JoSaV} and Liu \cite{Liu2015JoSaV} when the external flow Mach number $M=0$ is performed; the results are shown in Fig.\ref{fig:valid3zhou}. The derived poroelastic field expressions are subsequently verified by these two cases. 

\subsubsection{Oblique incident STL validation with the FEM results}\label{subsec:fem_validation}

\begin{figure}[!htb]
     \centering
     \includegraphics[width=.8\textwidth]{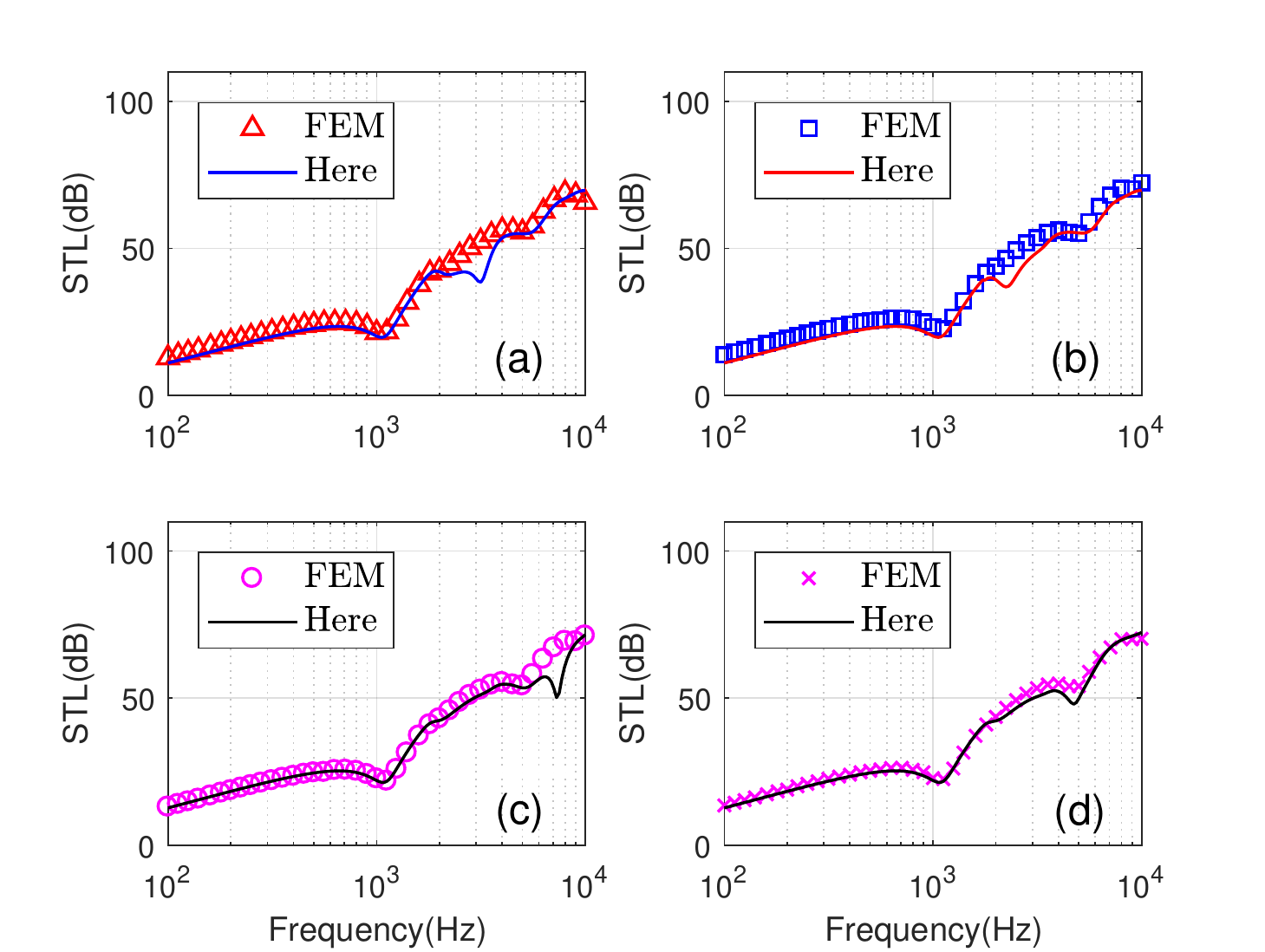}
     \caption{Validation of the 3D oblique incidence BB boundary case: (a) $\varphi_1=\pi/4$, $\theta_1=\pi/4$; (b) $\varphi_1=\pi/4$, $\theta_1=\pi/3$; (c) $\varphi_1=\pi/3$, $\theta_1=\pi/4$; (d) $\varphi_1=\pi/3$, $\theta_1=\pi/3$}\label{fig:validfem3dbb}
\end{figure}

\begin{figure}[!htb]
     \centering
     \includegraphics[width=.8\textwidth]{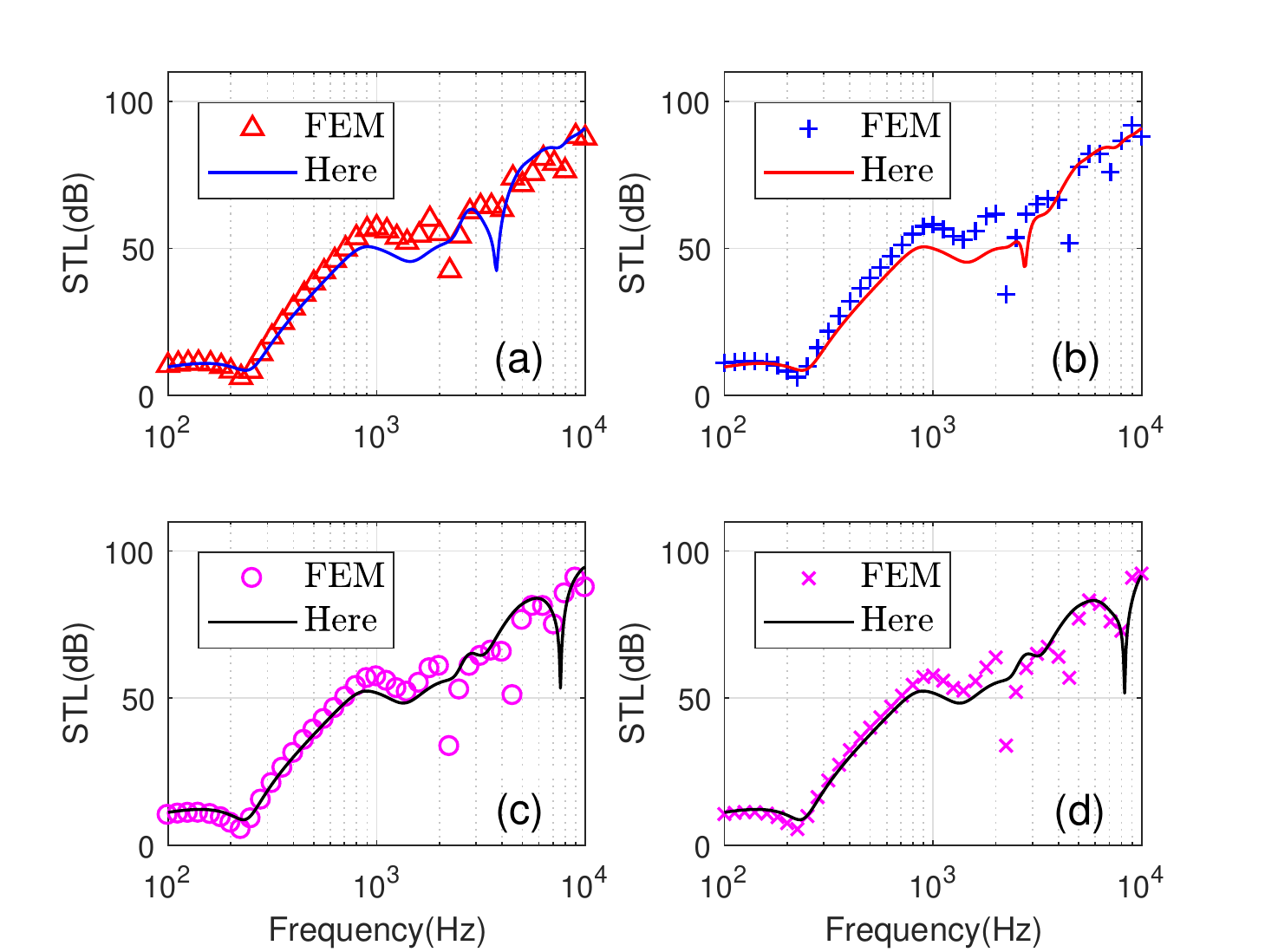}
     \caption{Validation of the 3D oblique incidence BU boundary case: (a) $\varphi_1=\pi/4$, $\theta_1=\pi/4$; (b) $\varphi_1=\pi/4$, $\theta_1=\pi/3$; (c) $\varphi_1=\pi/3$, $\theta_1=\pi/4$; (d) $\varphi_1=\pi/3$, $\theta_1=\pi/3$}\label{fig:validfem3dbu}
\end{figure}

\begin{figure}[!htb]
     \centering
     \includegraphics[width=.8\textwidth]{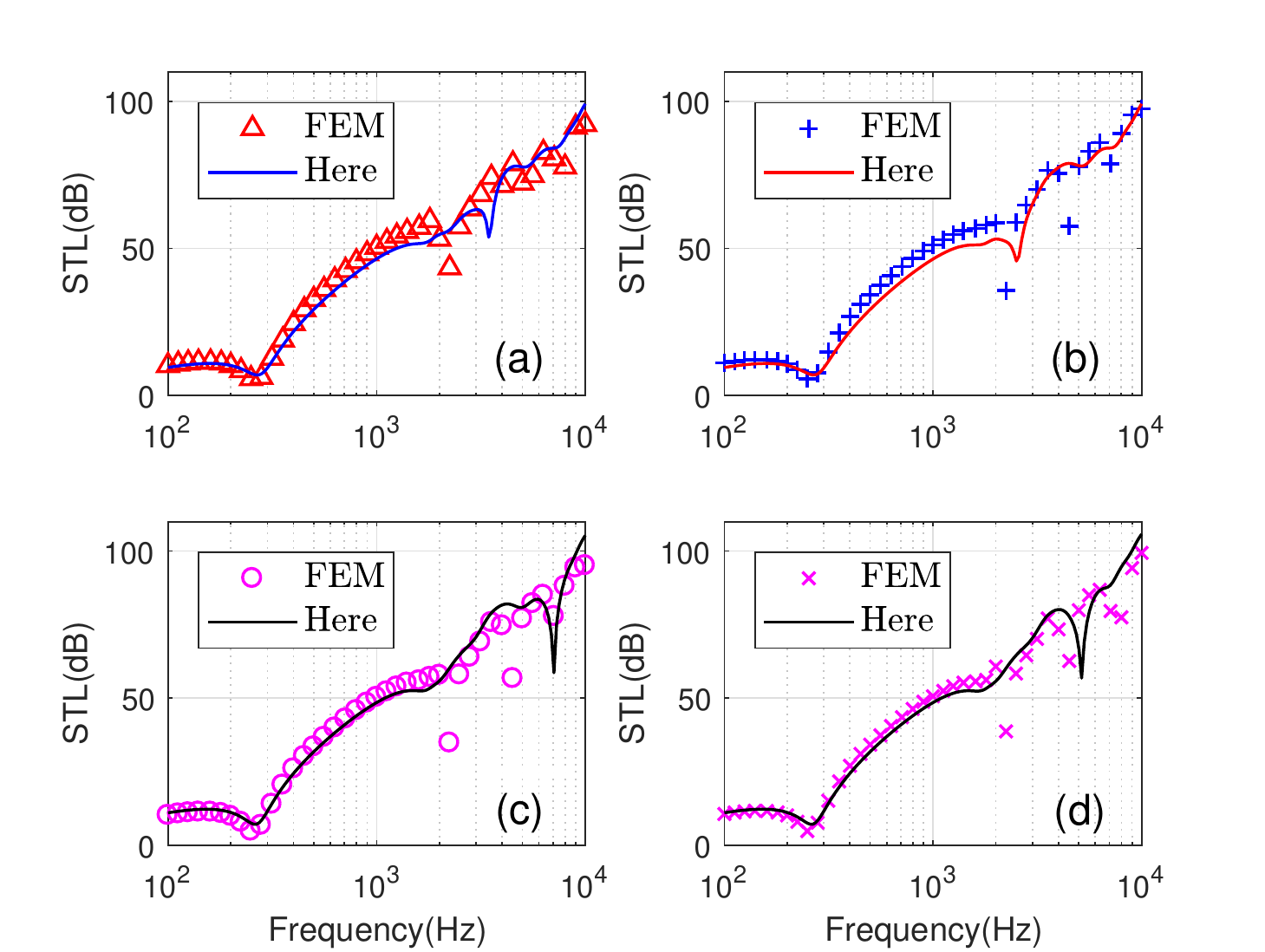}
     \caption{Validation of the 3D oblique incidence UU boundary case: (a) $\varphi_1=\pi/4$, $\theta_1=\pi/4$; (b) $\varphi_1=\pi/4$, $\theta_1=\pi/3$; (c) $\varphi_1=\pi/3$, $\theta_1=\pi/4$; (d) $\varphi_1=\pi/3$, $\theta_1=\pi/3$}\label{fig:validfem3duu}
\end{figure}

Several oblique incident cases are compared, as the random STL is clearly related to the oblique incident case according to Eq.(\ref{eq:randomstl}). The oblique incident STL are used in Figs.\ref{fig:validfem3dbb}-\ref{fig:validfem3duu} for the vertical axis values.

\begin{figure}[!htb]
     \centering
     \includegraphics[width=.45\textwidth]{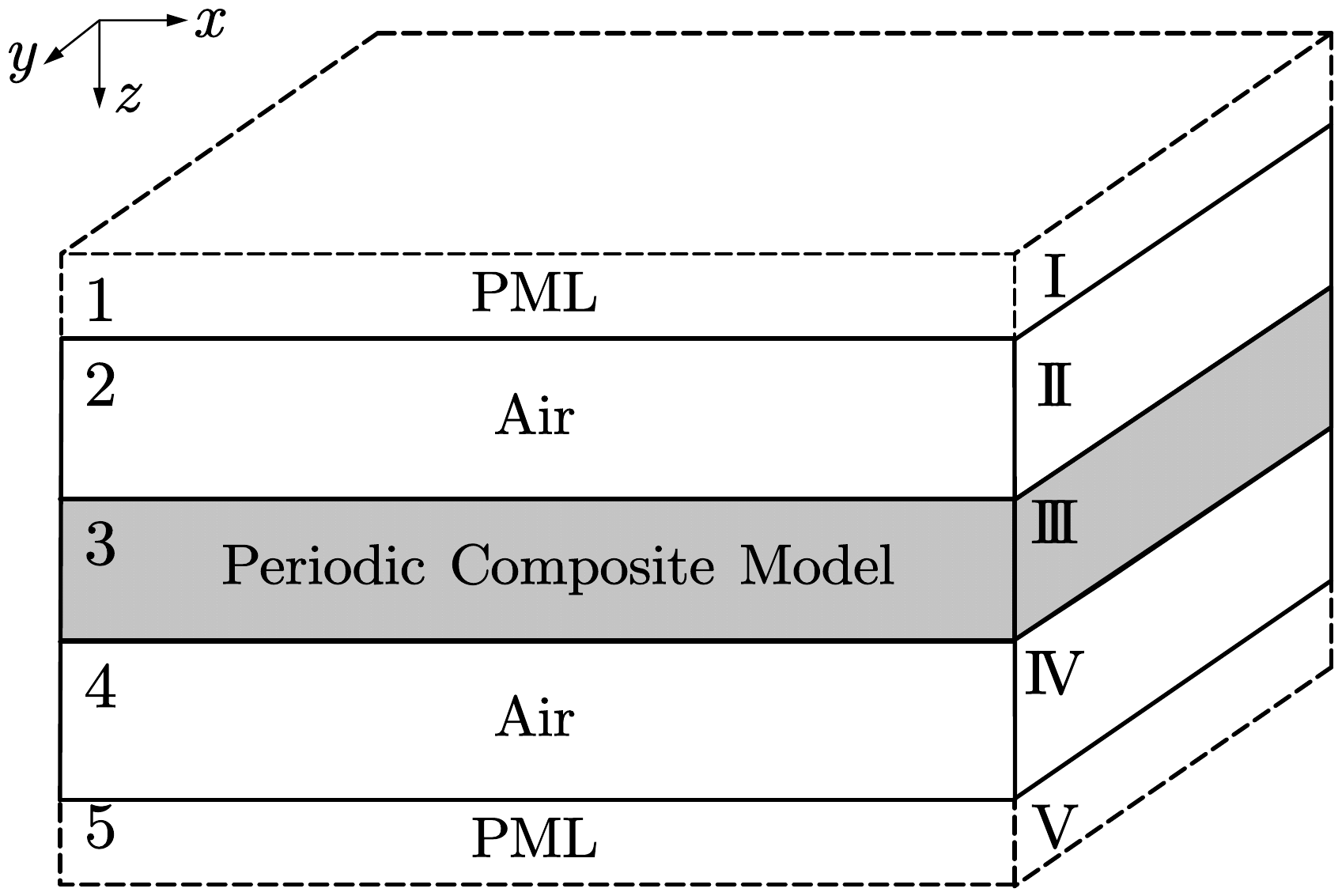}
     \caption{Schematic diagram of the FEM model: the periodic BC is applied on the surfaces vertical to x (the surfaces I--V etc.); the symmetric BC is applied on the surfaces vertical to y (the surfaces 1--5 etc.)}\label{fig:valid3fem}
\end{figure}

The FEM models are formed by the mixed displacement--pressure form of the Biot--Allard equations \cite{Allard2009}; the viscous and thermal characteristic lengths required are obtained using an equivalent relationship \cite{Allard1992TJotASoA,Allard2009} $\Lambda=(8\mu_f\epsilon'/\sigma\phi)^{1/2}$, $\Lambda'=2\Lambda$; here, $\mu_f$ is the dynamic viscosity of the fluid in the pores. The perfectly matched layer (PML) and periodic boundary conditions (periodic BC) are used. The symmetric boundary conditions (symmetric BC) for the y coordinate are also used, as the structure is infinite in the y direction. A schematic description of the FEM configuration is shown in Fig.\ref{fig:valid3fem}. The FEM calculations are performed using COMSOL. 

The results are given in Figs.\ref{fig:validfem3dbb}-\ref{fig:validfem3duu} for the BB, BU, and UU cases, respectively. The overall consistency and critical local differences with the FEM results are both satisfactory. In all oblique incidence cases, on average, the STL differences are less than 2 dB (absolute value), while the relative differences are all no more than 10\% of the FEM results except for a few asynchronous extrema. This is another evidence of the feasibility of the proposed method herein.

The time used in the 3D oblique incidence calculations are listed in Table \ref{tab:theory_vs_fem}. The efficiency is pronounced, while the FEM is expensive owing to the refined model required by a large size change. However, the drawback should be noted, as the convergence verification can consume up to 69.76\%, 73.34\% and 75.45\% of the total calculation time on average; for the 3D random incidence cases (the following parameter analyses), more than 80\% of the total calculation time could be used; therefore, an improvement is required. A related study is in progress.

\begin{table}[!ht]
\centering
\caption{Time used by the oblique incidence cases: CC.T is the convergence check time, and C.T is the calculation time. A periodic span along x is used in the FE Model; there are 310228, 202882, and 270319 unstructured 3D elements used for the BB, BU, and UU cases, respectively}\label{tab:theory_vs_fem}
\resizebox{\textwidth}{16mm}{
\begin{tabular}{|c|c|c|c|c|c|c|c|c|c|}
\hline
 & \multicolumn{3}{|c|}{BB Case (s)} & \multicolumn{3}{|c|}{BU Case (s)} & \multicolumn{3}{|c|}{UU Case (s)} 
 \\
 ($\varphi_1$,$\theta_1$) & FEM & CC.T & C.T & FEM &  CC.T & C.T & FEM &  CC.T & C.T 
\\
\hline
 ($\pi/4,\pi/4$) & 39482 & 4.522 & 2.134 & 14132 & 4.054 & 2.152 & 10593 & 3.610 & 1.645
\\
 ($\pi/4,\pi/3$) & 65396 & 3.305 & 1.554 & 13690 & 4.003 & 1.535 & 10461 & 3.173 & 1.088 
\\
 ($\pi/3,\pi/4$) & 53677 & 3.736 & 1.526 & 13940 & 4.945 & 1.537 & 10638 & 4.018 & 1.118 
\\
 ($\pi/3,\pi/3$) & 40299 & 3.980 & 1.544 & 15151 & 5.209 & 1.348 & 10823 & 4.507 & 1.098 
 \\
\hline
\end{tabular}}
\end{table}

\subsection{Influence of the rib reinforcement model on the {\rm STL}}

\begin{figure}[!htb]
     \centering
     \includegraphics[width=.8\textwidth]{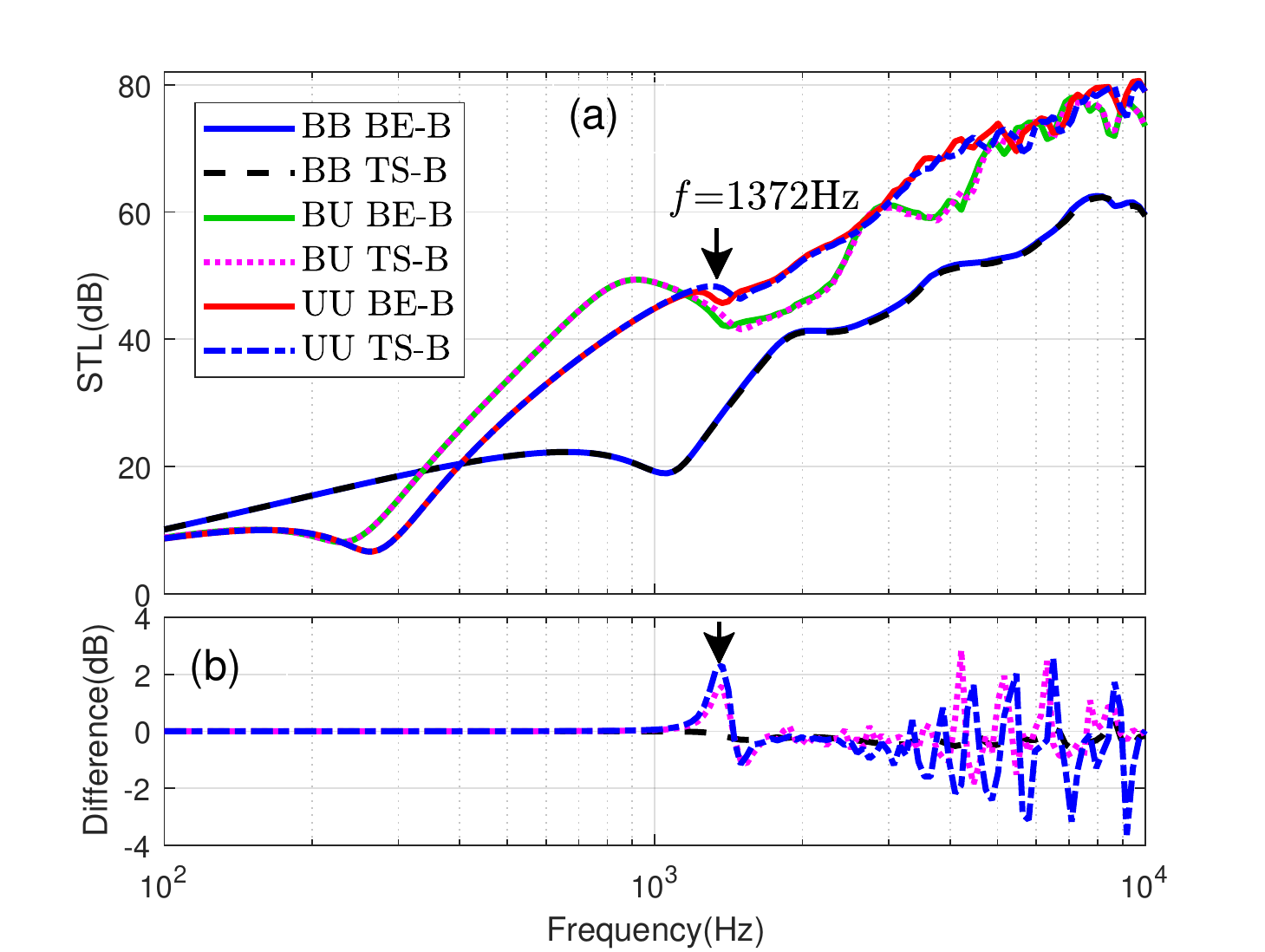}
     \caption{Results of different beam models (a) the STL of different cases; (b) difference between beam models in each boundary condition. The legends of (b) are the same as (a)}\label{fig:m3dbeammodel}
\end{figure}

The beam models mentioned above are used. 
The STL and its difference defined as the Timoshenko beam case minus the Bernoulli--Euler beam case for different boundary conditions are shown in Fig.\ref{fig:m3dbeammodel}.

\begin{figure}[!htb]
     \centering
     \includegraphics[width=.8\textwidth]{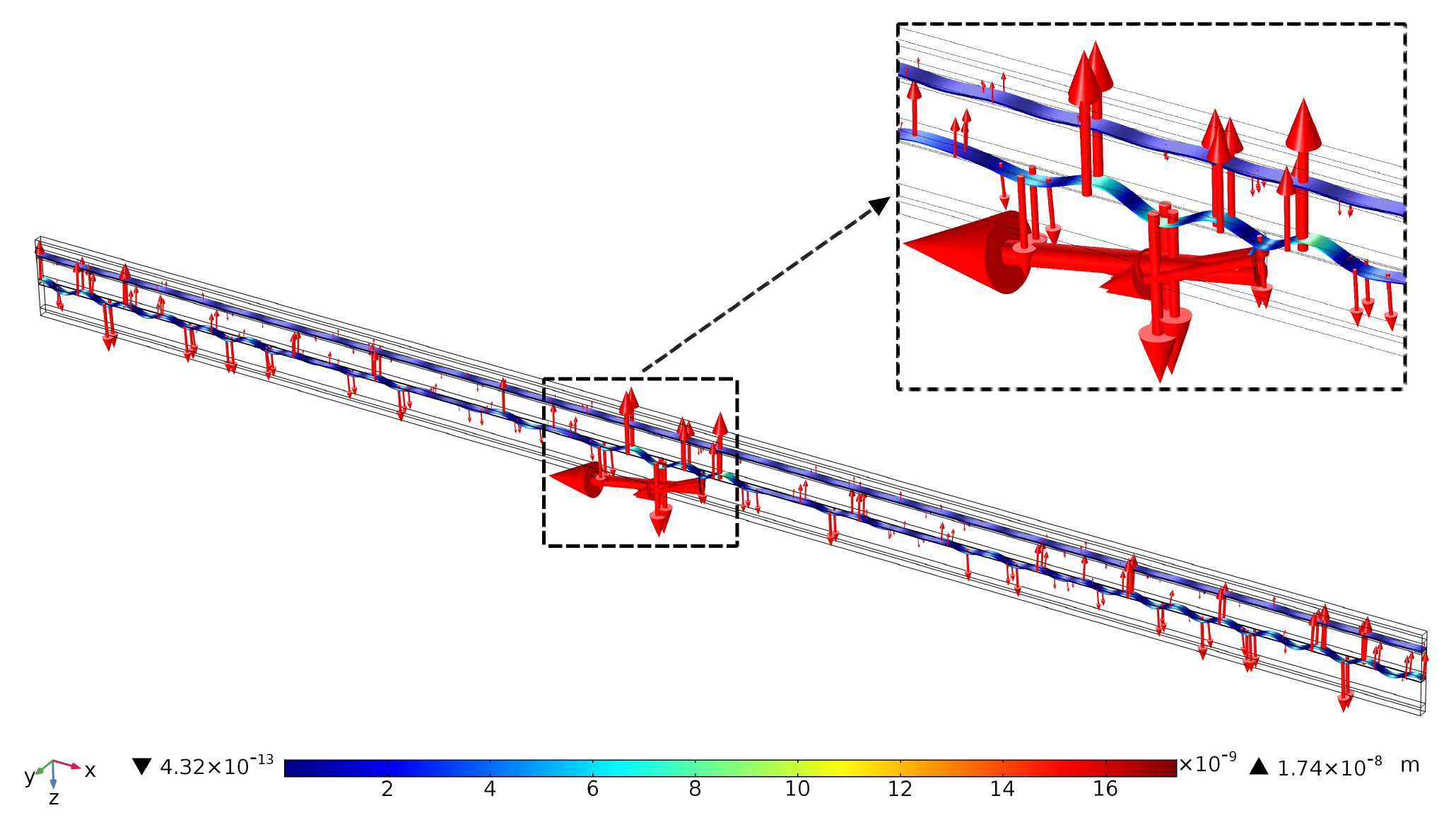}
     \caption{UU case, eigenfrequency = 1375 Hz. surface contour: the magnitude of total displacement; arrow: the magnitude and direction of displacement vector ; dotted line: the mode shape around ribs}\label{fig:bm_uu_1375}
\end{figure}

An overall consistency is found for the BB case, while approximately 2 dB differences (under most frequencies) are found for the BU and UU cases.
 It is confirmed by the FEM results (the configuration in section \ref{subsec:fem_validation} is used hereinafter) that the combined flexural and torsional modes of the ribs around 1372 Hz appeared; one of these modes (UU case, eigenfrequency = 1375 Hz) is shown in Fig.\ref{fig:bm_uu_1375}. Owing to the consistency of the two beam models, the Bernoulli--Euler beam model is preferable. It is used in the following.

\subsection{Influence of the torsion motion on the {\rm STL}}

\begin{figure}[!htb]
     \centering
     \includegraphics[width=.8\textwidth]{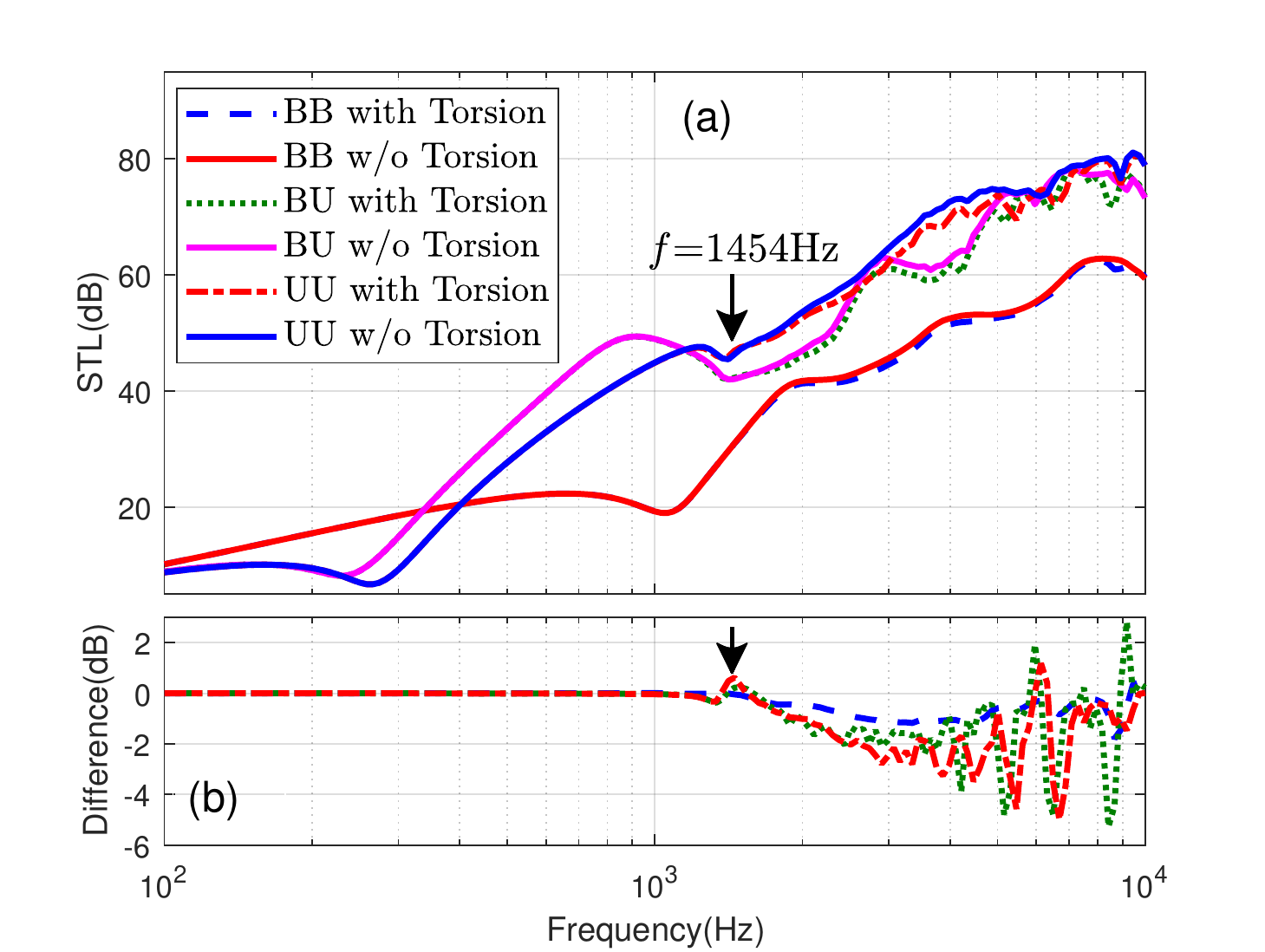}
     \caption{Results of torsion motion (a) STL when torsion is present or absent; (b) STL decrease when torsion is considered. The legends of (b) are the same as (a)}\label{fig:torsion}
\end{figure}
 
\begin{figure}[!htb]
     \centering
     \includegraphics[width=.8\textwidth]{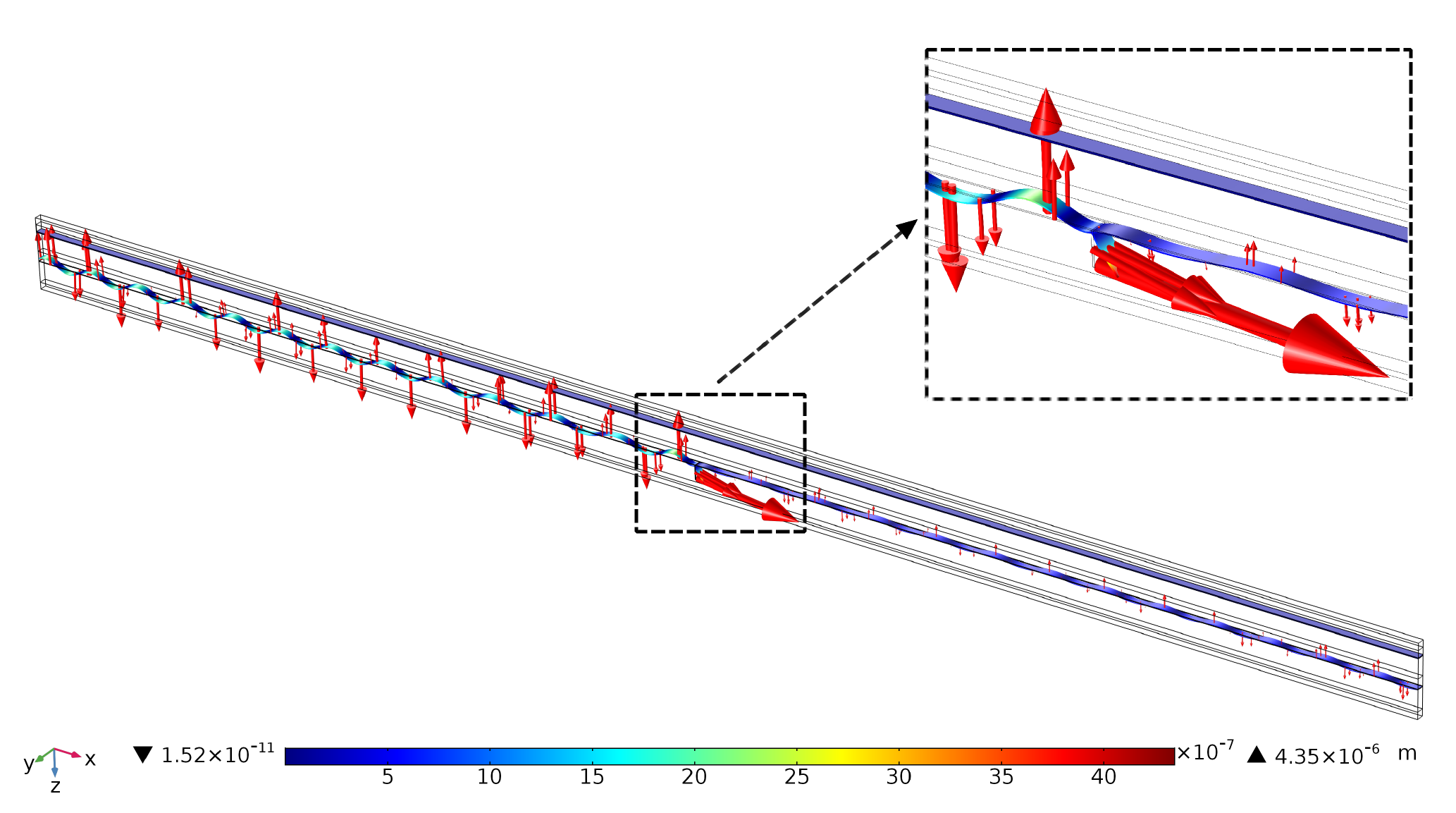}
     \caption{BU case, eigenfrequency = 1455 Hz: the torsional modes of ribs. surface contour: the magnitude of total displacement; arrow: the magnitude and direction of displacement vector; dotted line: the mode shape around ribs}\label{fig:tor_bu_1454}
\end{figure}

The influence of torsion is discussed, although the transverse motion is more important in the acoustics; the detailed results are shown in Fig.\ref{fig:torsion}. The STL decrease emerges and can be larger than 5 dB when the frequency exceeds 1454 Hz (indicated in Fig.\ref{fig:torsion}). This is due to the emergence of the torsional mode of ribs (confirmed by the FEM results, as shown in Fig.\ref{fig:tor_bu_1454}). 
Owing to the energy consumption by the torsional mode, the STL increases temporarily around the eigenfrequency and deteriorates when it is further from it. The torsion motion, with its influence on the STL, should not be neglected.

\subsection{Influence of the rib spacing on the {\rm STL}}
 
The results when the rib spacing $l_x$ varies among $l_x/d= 20,50,150$ versus the unribbed case are discussed. As the results of the BB case are analogous to the BU and UU cases, the details are provided in the supplementary material.
 
\begin{figure}[!htb]
     \centering
     \includegraphics[width=.8\textwidth]{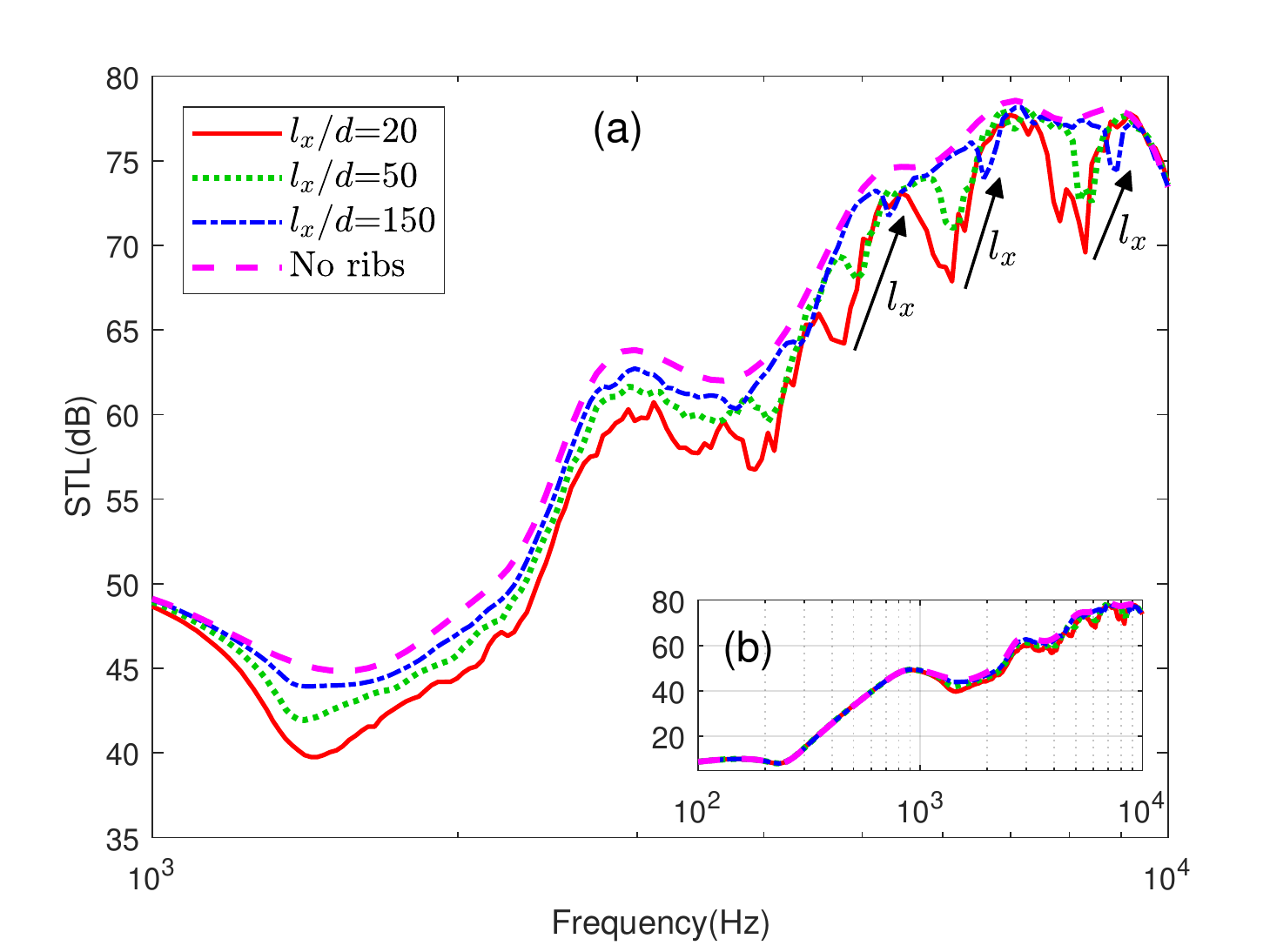}
     \caption{Influence of $l_x$ in the BU case : (a) STL at [1 kHz,10 kHz]; (b) overall STL trend. The black arrow indicates the direction $l_x$ increases}\label{fig:lxbu}
\end{figure}

\begin{figure}[!htb]
     \centering
     \includegraphics[width=.8\textwidth]{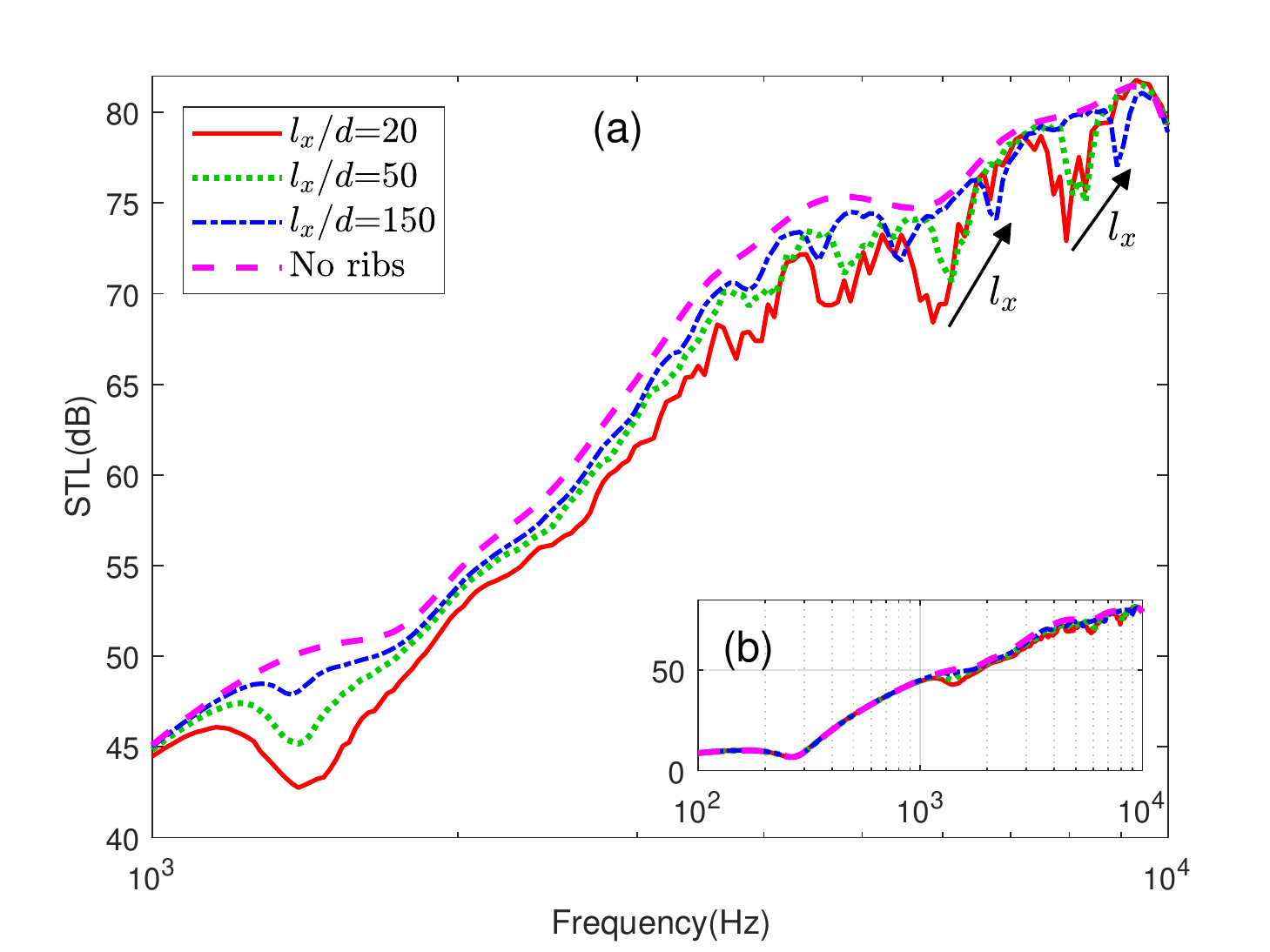}
     \caption{Influence of $l_x$ in the UU case : (a) STL at [1 kHz,10 kHz]; (b) overall STL trend. The black arrow indicates the direction $l_x$ increases}\label{fig:lxuu}
\end{figure}

In all three cases, the STL is positively related to $l_x$ in general, while degradation occurs compared to the unribbed one. The overall trends are analogous to the unribbed results \cite{Bolton1996JoSaV,Zhou2013JoSaV,Liu2015JoSaV}. As $l_x$ decreases, more fluctuations emerge and the trough frequencies shift to the lower ones (indicated in Figs.\ref{fig:lxbu}-\ref{fig:lxuu}). 
In fact, to avoid the wave interference between the ribs, $l_x$ should be far greater than the coincidence wavelength $\lambda_p$ of the face panels, which is \cite{Junger1986}
\begin{equation}\label{eq:lambda_p}
 \lambda_p=c_0/f_c,\ f_c=c_0^2/2\pi\sqrt{\rho_p h_i /D}
\end{equation}
 Here, $c_0$ is the sound velocity of the adjacent medium, $i=1,2$ for the incident and transmitted side panels respectively; $\lambda_p=36,22$ (mm), respectively. When $l_x$ is not sufficiently large, the interference can be complex and fluctuations emerge. 

In general, the STL is positively related to $l_x$ even while ribs lower to it are present, although a direct transfer path is absent. This is due to the increase in the flexural wave speed of the composite structure, i.e., the influence of stiffness increase is stronger than that of mass increase \cite{Hambric2016} when the ribs are present. To obtain a better sound insulation performance, the ribs should be removed or $l_x$ should be increased at the least. Further investigations on the dispersion relation are required to present a quantitative conclusion.

\subsection{Influence of area moment of inertia on the {\rm STL}}
 
To exclude the influence of mass change, the cross-section area of ribs $A=t_x h_x$ is maintained constant; subsequently, the area moment of inertia $I$ is inversely proportional to $t_x/h_x$. Therefore, the investigation is performed with $t_x/h_x=1/3,1/5,1/10,1/20$. 

\begin{figure}[!htb]
     \centering
     \includegraphics[width=.8\textwidth]{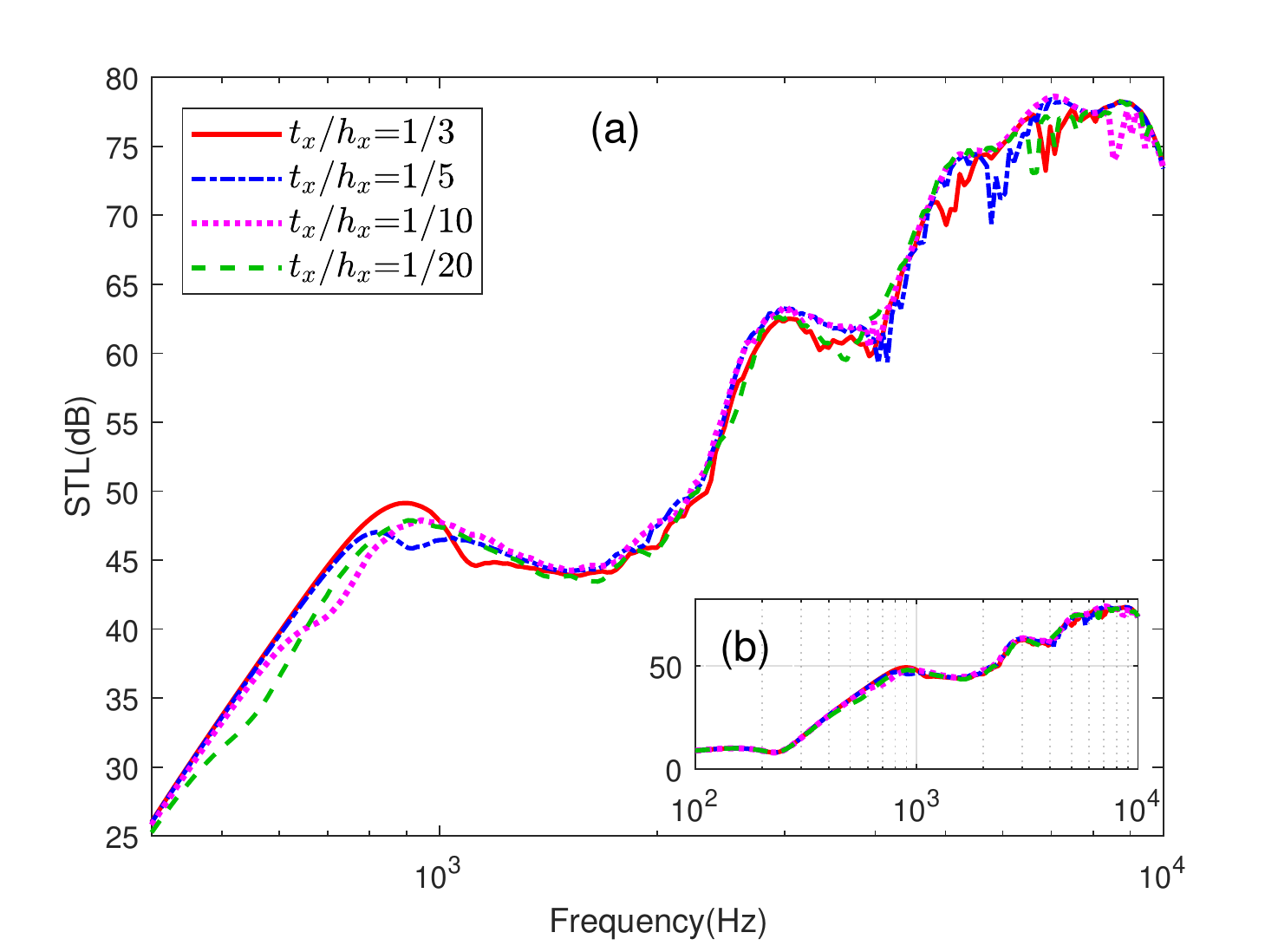}
     \caption{Influence of $I$ in the BU case : (a) STL at [400 Hz,10 kHz]; (b) overall STL trend}\label{fig:momentbu}
\end{figure}

\begin{figure}[!htb]
     \centering
     \includegraphics[width=.8\textwidth]{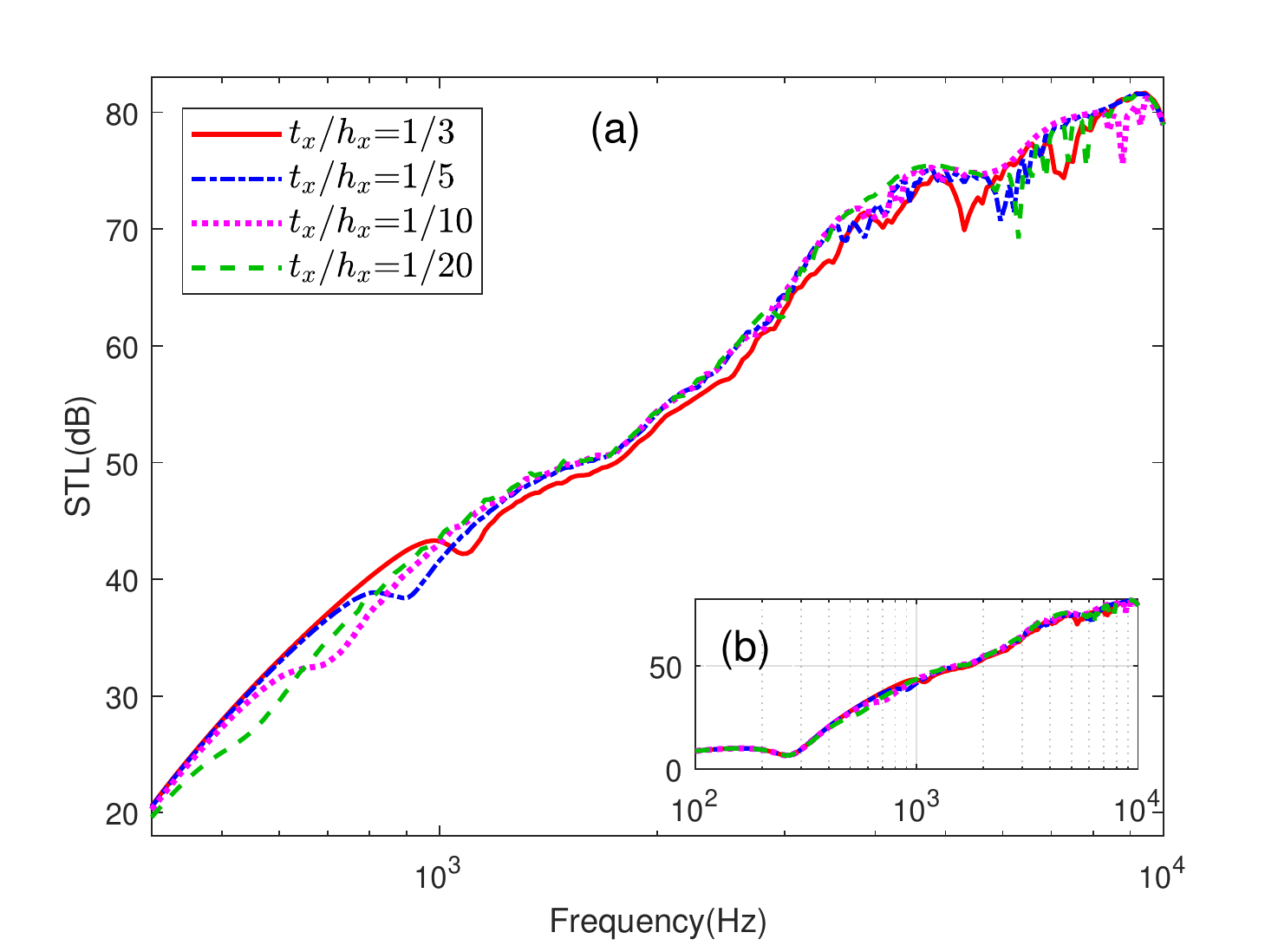}
     \caption{Influence of $I$ in the UU case : (a) STL at [400 Hz,10 kHz]; (b) overall STL trend}\label{fig:momentuu}
\end{figure}
 
In the BB case, as the structural connections are strong, the influence of $I$ is found to be weak; therefore, the results are presented in the supplementary material. As shown in Figs.\ref{fig:momentbu}--\ref{fig:momentuu}, the STL decreases with the increase in $I$ (as $t_x/h_x$ decreases) under a relatively low frequency range (i.e., between 100 Hz and 1 kHz); this is due to the increase in the phase velocity $v_b$ of the ribs \cite{Junger1986}
\begin{equation}
v_b=\left({{E I \omega^2}/{\rho A}} \right)^{{1}/{4}}
\end{equation}
which is proportional to the group velocity in this study; thus, a better energy transmission and lower insulation are anticipated. In the higher frequency range, the STL curve is analogous to the unribbed case \cite{Bolton1996JoSaV,Zhou2013JoSaV}; it is the superposition of the unribbed STL and the fluctuations caused by the ribs. 

In general, the influence of $I$ occurs in the low-frequency range, within which the STL is negatively related to $I$. To obtain a better sound insulation in the low-frequency range, a smaller area moment of inertia is preferred for similar structures.

\section{Conclusions}\label{stn:conclusion}
We herein proposed a one-dimensional periodic composite structure: a periodically rib-stiffened double panel with porous lining, to investigate the effects of combining a periodic structure and poroelastic problem. This periodic poroelastic problem was studied in terms of its sound insulation using a semi-analytical model developed based on the Biot theory and SHS. 

We found that the sound insulation was insensitive to the rib reinforcement model in these structures, while the torsional motion, the rib spacing, and the area moment of inertia were important; however, their influences were exhibited in different frequency ranges. The presence of periodic ribs was found to lower the overall sound insulation, although a direct transfer path was absent. This is because the flexural wave speed of the periodic composite structure is increased, and thus its sound insulation is not improved. Additional research should focus on the dispersion relation to quantitatively understand the combined effects. The convergence efficiency is also important and requires further investigation.

Despite the unexpected results from the model, the method proposed for the periodic poroelastic problem, which is efficient with considerable accuracy, can be promising in broadband sound modulation.

\section*{Acknowledgements}
This work is supported by the National Natural Science Foundation of China (NSFC) No.11572137. The authors wish to thank the anonymous researchers and reviewers 
for their warm-hearted suggestions and comments. 

\appendix

\section{The elements of the porous field variable coefficient matrix}\label{app:porousexpression}
To be brief, here $k_x^{m}$,\ $k_y^{m}$,\ $k_{1,z}^m$,\ $k_{2,z}^m$ and $k_{3,z}^m$ are replaced by $\alpha_m$,\ $\beta$,\ $\gamma_{1,m}$,\ $\gamma_{2,m}$ and $\gamma_{3,m}$ respectively, the elements of the coefficient matrix ${\bf Y}_m$ are
\begin{gather}
Y_m(1,1)=Y_m(1,2)=\frac{{\rm j}\alpha_m}{k_1^2},\ Y_m(1,3)=Y_m(1,4)=\frac{{\rm j}\alpha_m}{k_2^2}
\notag\\
Y_m(1,5)=-\frac{{\rm j}}{k_3^2\gamma_{3,m}}[(\gamma_{3,m}^2+\beta^2){\rm cos}\theta-\alpha_m\beta{\rm sin}\theta],\ Y_m(1,6)=-Y_m(1,5)
\notag\\
Y_m(2,1)=Y_m(2,2)=\frac{{\rm j}\beta}{k_1^2},\ Y_m(2,3)=Y_m(2,4)=\frac{{\rm j}\beta}{k_2^2}
\notag\\
Y_m(2,5)=-\frac{{\rm j}}{k_3^2\gamma_{3,m}}[(\gamma_{3,m}^2+\alpha_m^2){\rm sin}\theta-\alpha_m\beta{\rm cos}\theta],\ Y_m(2,6)=-Y_m(2,5)
\notag\\
Y_m(3,1)=\frac{{\rm j}\gamma_{1,m}}{k_1^2},\ Y_m(3,2)=-Y_m(3,1)
\notag\\
Y_m(3,3)=\frac{{\rm j}\gamma_{2,m}}{k_2^2},\ Y_m(3,4)=-Y_m(3,3)
\notag\\
Y_m(3,5)=\frac{{\rm j}}{k_3^2}(\alpha_m{\rm cos}\theta+\beta{\rm sin}\theta),\ Y_m(3,6)=Y_m(3,5)
\notag\\
Y_m(4,1)=Y_m(4,2)=\frac{{\rm j} b_1\alpha_m}{k_1^2},\ Y_m(4,3)=Y_m(4,4)=\frac{{\rm j} b_2\alpha_m}{k_2^2}
\notag\\
Y_m(4,5)=-\frac{{\rm j} g}{k_3^2\gamma_{3,m}}[(\gamma_{3,m}^2+\beta^2){\rm cos}\theta-\alpha_m\beta{\rm sin}\theta],\ Y_m(4,6)=-Y_m(4,5)
\notag\\
Y_m(5,1)=Y_m(5,2)=\frac{{\rm j} b_1\beta}{k_1^2},\ Y_m(5,3)=Y_m(5,4)=\frac{{\rm j} b_2\beta}{k_2^2}
\notag\\
Y_m(5,5)=-\frac{{\rm j} g}{k_3^2\gamma_{3,m}}[(\gamma_{3,m}^2+\alpha_m^2){\rm sin}\theta-\alpha_m\beta{\rm cos}\theta],\ Y_m(5,6)=-Y_m(5,5)
\notag\\
Y_m(6,1)=\frac{{\rm j} b_1\gamma_{1,m}}{k_1^2},\ Y_m(6,2)=-Y_m(6,1)
\notag\\
Y_m(6,3)=\frac{{\rm j} b_2\gamma_{2,m}}{k_2^2},\ Y_m(6,4)=-Y_m(6,3)
\notag\\
Y_m(6,5)=\frac{{\rm j} g}{k_3^2}(\alpha_m{\rm cos}\theta+\beta{\rm sin}\theta),\ Y_m(6,6)=Y_m(6,5)
\notag
\end{gather}
where $b_1=a_1-a_2 k_1^2$, $b_2=a_1-a_2 k_2^2$; while 
\begin{equation}
\begin{cases} 
{\rm cos}\theta=-\alpha_m/\sqrt{\alpha_m^2+\beta^2},\ {\rm sin}\theta=-\beta/\sqrt{\alpha_m^2+\beta^2}, & \mbox{if }\alpha_m^2+\beta^2\ne\mbox{0} \\
{\rm cos}\theta={\rm Inf},\ {\rm sin}\theta={\rm Inf}, & \mbox{if }\alpha_m^2+\beta^2=\mbox{0}
\end{cases}\notag
\end{equation}
Here, ${\rm Inf}$ can be chosen as a large number (e.g., ${\rm Inf}=10^{500}$), as at the normal incidence ($\alpha_m^2+\beta^2=0$) no shear wave is excited \cite{Allard2009}, $C_5^m=C_6^m=0$ for all $m\in[-\hat{m},\hat{m}]$.

\section{Derivation of the double summation identity Eq.(\ref{eq:identity})}\label{app:identityproof}
The basic idea to prove the identity is to use ${\rm e}^{-{\rm j} {k}_{x}^{m}x } {\rm e}^{{\rm j} 2\pi n x/l_x} = {\rm e}^{-{\rm j} {k}_{x}^{(m-n)}x}$, as $k_{x}^{m}=k_x+{2m\pi}/{l_x}$. Subsequently, the left-hand side of the identity becomes
\begin{equation}\label{eq:identityp1}
\sum_{n=-\infty}^{+\infty}\left( \cdots +W_{-1}\ {\rm e}^{-{\rm j} {k}_{x}^{-1-n}x}+W_{0}\ {\rm e}^{-{\rm j} {k}_{x}^{-n}x}+W_{1}\ {\rm e}^{-{\rm j} {k}_{x}^{1-n}x}+\cdots \right)
\end{equation}

As the summation index $n\in[-\infty,+\infty]$ in Eq.(\ref{eq:identityp1}), if the summation is expanded and the index $n$ loops over $n-1$ to $n+1$, subsequently Eq.(\ref{eq:identityp1}) becomes
\begin{equation}\label{eq:identityp2}
\begin{split}
\Big\{\cdots&+\left( \cdots +W_{-1} \ {\rm e}^{-{\rm j} {k}_{x}^{-n}x}+W_{0} \ {\rm e}^{-{\rm j} {k}_{x}^{-n+1}x}+W_{1} \ {\rm e}^{-{\rm j} {k}_{x}^{-n+2}x}+\cdots \right)+\\
&+\left( \cdots +W_{-1} \ {\rm e}^{-{\rm j} {k}_{x}^{-n-1}x}+W_{0} \ {\rm e}^{-{\rm j} {k}_{x}^{-n}x}+W_{1} \ {\rm e}^{-{\rm j} {k}_{x}^{-n+1}x}+\cdots \right)+\\
&+\left( \cdots +W_{-1} \ {\rm e}^{-{\rm j} {k}_{x}^{-n-2}x}+W_{0} \ {\rm e}^{-{\rm j} {k}_{x}^{-n-1}x}+W_{1} \ {\rm e}^{-{\rm j} {k}_{x}^{-n}x}+\cdots \right)+\cdots\Big\}
\end{split}
\end{equation}

If the terms with ${\rm e}^{-{\rm j} {k}_{x}^{-n}x}$,\ ${\rm e}^{-{\rm j} {k}_{x}^{-n+1}x}$,\ ${\rm e}^{-{\rm j} {k}_{x}^{-n+2}x}$... in Eq.(\ref{eq:identityp2}) were collected individually, Eq.(\ref{eq:identityp2}) becomes
\begin{equation}\label{eq:identityp3}
\begin{split}
\Big\{\cdots&+{\rm e}^{-{\rm j} {k}_{x}^{-n}x}\left(\cdots+W_{-1}+W_{0}+W_{1}+\cdots \right)+\cdots\\
&+{\rm e}^{-{\rm j} {k}_{x}^{-n+1}x}\left(\cdots+W_{0}+W_{1}+W_{2}+\cdots \right)+\cdots\\
&+{\rm e}^{-{\rm j} {k}_{x}^{-n+2}x}\left(\cdots+W_{1}+W_{2}+W_{3}+\cdots \right)+\cdots\Big\}
\end{split}
\end{equation}
Subsequently, the double summation identity Eq.(\ref{eq:identity}) is obtained
\begin{equation}
\sum_{n=-\infty}^{+\infty}\sum_{m=-\infty}^{+\infty}W_m {\rm e}^{-{\rm j} {k}_{x}^{m}x} {\rm e}^{{\rm j} 2\pi n x/l_x}=\sum_{n=-\infty}^{+\infty}W_n \sum_{m=-\infty}^{+\infty} {\rm e}^{-{\rm j} {k}_{x}^{m}x}
\end{equation}

\section{The elements of the matrices in Eq.(\ref{eq:sysmatrix})}\label{app:matrixelements}
To be brief, here $k_x^{m}$,\ $k_y^{m}$,\ $k_{i,z}^m$,\ $k_{1,z}^m$,\ $k_{2,z}^m$,\ $k_{3,z}^m$ and $k_{t,z}^m$ are replaced by $\alpha_m$,\ $\beta$,\ $\gamma_{i,m}$,\ $\gamma_{1,m}$,\ $\gamma_{2,m}$,\ $\gamma_{3,m}$ and $\gamma_{t,m}$ respectively, and denote $L_1=h_1+h_p$,\ $L_2=h_1+h_p+h_2/2$,\ $L_3=h_1+h_p+h_2$, subsequently the elements of matrix ${\bf A}_m$ can be deduced as

\begin{gather}
A_m(1,9)={\rm j} \omega,\ A_m(1,13)={\rm j} \gamma_{i,m}
\notag\\
A_m(2,1)=2\frac{N \gamma_{1,m} \alpha_m}{k_1^2}{\rm e}^{-{\rm j} h_1 \gamma_{1,m}/2},\ A_m(2,2)=-2\frac{N \gamma_{1,m} \alpha_m}{k_1^2}{\rm e}^{{\rm j} h_1 \gamma_{1,m}/2}
\notag\\
A_m(2,3)=2\frac{N \gamma_{2,m} \alpha_m}{k_2^2}{\rm e}^{-{\rm j} h_1 \gamma_{2,m}/2},\ A_m(2,4)=-2\frac{N \gamma_{2,m} \alpha_m}{k_2^2}{\rm e}^{{\rm j} h_1 \gamma_{2,m}/2}
\notag\\
A_m(2,5)=-\frac{N {\rm e}^{-{\rm j} h_1 \gamma_{3,m}/2}}{k_3^2}\left[(\beta^2+\gamma_{3,m}^2-\alpha_m^2){\rm cos}\theta-2\beta\alpha_m{\rm sin}\theta\right]
\notag\\
A_m(2,6)=-\frac{N {\rm e}^{{\rm j} h_1 \gamma_{3,m}/2}}{k_3^2}\left[(\beta^2+\gamma_{3,m}^2-\alpha_m^2){\rm cos}\theta-2\beta\alpha_m{\rm sin}\theta\right]
\notag\\
A_m(2,7)=-\frac{1}{2}(1-\nu_p)D_{p1}\beta^2+m_{s1}\omega^2-D_{p1}\alpha_m^2
\notag\\
A_m(2,8)=-\frac{1}{2}(1+\nu_p)D_{p1}\beta \alpha_m 
\notag\\
A_m(3,1)=[A+Q+b_1 E_f+2N\frac{\gamma_{1,m}^2}{k_1^2}-{\rm j} h_1 N\gamma_{1,m}\frac{\alpha_m^2+\beta^2}{k_1^2}]{\rm e}^{-{\rm j} h_1 \gamma_{1,m}/2}
\notag\\
A_m(3,2)=[A+Q+b_1 E_f+2N\frac{\gamma_{1,m}^2}{k_1^2}+{\rm j} h_1 N\gamma_{1,m}\frac{\alpha_m^2+\beta^2}{k_1^2}]{\rm e}^{{\rm j} h_1 \gamma_{1,m}/2}
\notag\\
A_m(3,3)=[A+Q+b_2 E_f+2N\frac{\gamma_{2,m}^2}{k_2^2}-{\rm j} h_1 N\gamma_{2,m}\frac{\alpha_m^2+\beta^2}{k_2^2}]{\rm e}^{-{\rm j} h_1 \gamma_{2,m}/2}
\notag\\
A_m(3,4)=[A+Q+b_2 E_f+2N\frac{\gamma_{2,m}^2}{k_2^2}+{\rm j} h_1 N\gamma_{2,m}\frac{\alpha_m^2+\beta^2}{k_2^2}]{\rm e}^{{\rm j} h_1 \gamma_{2,m}/2}
\notag\\
A_m(3,5)=-(\alpha_m{\rm cos}\theta +\beta {\rm sin}\theta)\left[(\alpha_m^2+\beta^2)h_1+4{\rm j}\gamma_{3,m}-h_1\gamma_{3,m}^2\right]\frac{{\rm j} N}{2 k_3^2}{\rm e}^{-{\rm j} h_1 \gamma_{3,m}/2}
\notag\\
A_m(3,6)=-(\alpha_m{\rm cos}\theta +\beta {\rm sin}\theta)\left[(\alpha_m^2+\beta^2)h_1-4{\rm j}\gamma_{3,m}-h_1\gamma_{3,m}^2\right]\frac{{\rm j} N}{2 k_3^2}{\rm e}^{{\rm j} h_1 \gamma_{3,m}/2}
\notag\\
A_m(3,9)=-D_1\alpha_m^4-2 D_1\alpha_m^2\beta^2-D_1\beta^4+m_{s1}\omega^2,\ A_m(3,13)={\rm j}\rho_i\omega {\rm e}^{{\rm j} h_1 \gamma_{i,m}/2} 
\notag\\
A_m(4,1)=\frac{{\rm j}\gamma_{1,m}}{k_1^2}{\rm e}^{-{\rm j} h_1 \gamma_{1,m}},\ A_m(4,2)=-\frac{{\rm j}\gamma_{1,m}}{k_1^2}{\rm e}^{{\rm j} h_1 \gamma_{1,m}}
\notag\\
A_m(4,3)=\frac{{\rm j}\gamma_{2,m}}{k_2^2}{\rm e}^{-{\rm j} h_1 \gamma_{2,m}},\ A_m(4,4)=-\frac{{\rm j}\gamma_{2,m}}{k_2^2}{\rm e}^{{\rm j} h_1 \gamma_{2,m}}
\notag\\
A_m(4,5)=(\alpha_m{\rm cos}\theta +\beta {\rm sin}\theta)\frac{{\rm j} {\rm e}^{-{\rm j} h_1 \gamma_{3,m}}}{k_3^2}
\notag\\
A_m(4,6)=(\alpha_m{\rm cos}\theta +\beta {\rm sin}\theta)\frac{{\rm j} {\rm e}^{{\rm j} h_1 \gamma_{3,m}}}{k_3^2}
\notag
\end{gather}
\begin{gather}
A_m(4,9)=-1,\ A_m(5,1)=\frac{{\rm j} b_1\gamma_{1,m}}{k_1^2}{\rm e}^{-{\rm j} h_1 \gamma_{1,m}}
\notag\\
A_m(5,2)=-\frac{{\rm j} b_1\gamma_{1,m}}{k_1^2}{\rm e}^{{\rm j} h_1 \gamma_{1,m}}
\notag\\
A_m(5,3)=\frac{{\rm j} b_2 \gamma_{2,m}}{k_2^2}{\rm e}^{-{\rm j} h_1 \gamma_{2,m}}
\notag\\
A_m(5,4)=-\frac{{\rm j} b_2 \gamma_{2,m}}{k_2^2}{\rm e}^{{\rm j} h_1 \gamma_{2,m}},\ A_m(5,5)=(\alpha_m{\rm cos}\theta +\beta {\rm sin}\theta)\frac{{\rm j} g {\rm e}^{-{\rm j} h_1 \gamma_{3,m}}}{k_3^2}
\notag\\
A_m(5,6)=(\alpha_m{\rm cos}\theta +\beta {\rm sin}\theta)\frac{{\rm j} g {\rm e}^{{\rm j} h_1 \gamma_{3,m}}}{k_3^2},\ A_m(5,9)=-1
\notag\\
\ A_m(6,1)=\frac{{\rm j}\alpha_m }{k_1^2}{\rm e}^{-{\rm j} h_1 \gamma_{1,m}},\ A_m(6,2)=\frac{{\rm j}\alpha_m }{k_1^2}{\rm e}^{{\rm j} h_1 \gamma_{1,m}}
\notag\\
A_m(6,3)=\frac{{\rm j}\alpha_m }{k_2^2}{\rm e}^{-{\rm j} h_1 \gamma_{2,m}},\ A_m(6,4)=\frac{{\rm j}\alpha_m }{k_2^2}{\rm e}^{{\rm j} h_1 \gamma_{2,m}} 
\notag\\
A_m(6,5)=-\frac{{\rm j} {\rm e}^{-{\rm j} h_1 \gamma_{3,m}}}{k_3^2\gamma_{3,m}}[\gamma_{3,m}^2{\rm cos}\theta+\beta(\beta {\rm cos}\theta-\alpha_m{\rm sin}\theta)] 
\notag\\
A_m(6,6)=\frac{{\rm j} {\rm e}^{{\rm j} h_1 \gamma_{3,m}}}{k_3^2\gamma_{3,m}}[\gamma_{3,m}^2{\rm cos}\theta+\beta(\beta {\rm cos}\theta-\alpha_m{\rm sin}\theta)] 
\notag\\
A_m(6,7)=-1,\ A_m(6,9)=-\frac{1}{2}{\rm j} h_1\alpha_m 
\notag\\
\ A_m(7,1)=\frac{{\rm j}\beta}{k_1^2}{\rm e}^{-{\rm j} h_1 \gamma_{1,m}},\ A_m(7,2)=\frac{{\rm j}\beta}{k_1^2}{\rm e}^{{\rm j} h_1 \gamma_{1,m}}
\notag\\
A_m(7,3)=\frac{{\rm j}\beta}{k_2^2}{\rm e}^{-{\rm j} h_1 \gamma_{2,m}},\ A_m(7,4)=\frac{{\rm j}\beta}{k_2^2}{\rm e}^{{\rm j} h_1 \gamma_{2,m}} 
\notag\\
A_m(7,5)=-\frac{{\rm j} {\rm e}^{-{\rm j} h_1 \gamma_{3,m}}}{k_3^2\gamma_{3,m}}[\gamma_{3,m}^2{\rm sin}\theta+\alpha_m(\alpha_m{\rm sin}\theta-\beta {\rm cos}\theta)] 
\notag\\
A_m(7,6)=\frac{{\rm j} {\rm e}^{{\rm j} h_1 \gamma_{3,m}}}{k_3^2\gamma_{3,m}}[\gamma_{3,m}^2{\rm sin}\theta+\alpha_m(\alpha_m{\rm sin}\theta-\beta {\rm cos}\theta)] 
\notag\\
A_m(7,8)=-1,\ A_m(7,9)=-\frac{1}{2}{\rm j} h_1\beta
\notag\\
A_m(8,1)=\frac{{\rm j}\gamma_{1,m}}{k_1^2}{\rm e}^{-{\rm j} L_1 \gamma_{1,m}},\ A_m(8,2)=-\frac{{\rm j}\gamma_{1,m}}{k_1^2}{\rm e}^{{\rm j} L_1 \gamma_{1,m}} 
\notag\\
A_m(8,3)=\frac{{\rm j}\gamma_{2,m}}{k_2^2}{\rm e}^{-{\rm j} L_1 \gamma_{2,m}},\ A_m(8,4)=-\frac{{\rm j}\gamma_{2,m}}{k_2^2}{\rm e}^{{\rm j} L_1 \gamma_{2,m}} 
\notag\\
A_m(8,5)=\frac{{\rm j} {\rm e}^{-{\rm j} L_1 \gamma_{3,m}}}{k_3^2}(\alpha_m{\rm cos}\theta+\beta {\rm sin}\theta)
\notag\\
A_m(8,6)=\frac{{\rm j} {\rm e}^{{\rm j} L_1 \gamma_{3,m}}}{k_3^2}(\alpha_m{\rm cos}\theta+\beta {\rm sin}\theta)
\notag\\
A_m(8,12)=-1,\ A_m(9,1)=\frac{{\rm j} b_1\gamma_{1,m}}{k_1^2}{\rm e}^{-{\rm j} L_1 \gamma_{1,m}} 
\notag
\end{gather}
\begin{gather}
A_m(9,2)=-\frac{{\rm j} b_1\gamma_{1,m}}{k_1^2}{\rm e}^{{\rm j} L_1 \gamma_{1,m}},\ A_m(9,3)=\frac{{\rm j} b_2\gamma_{2,m}}{k_2^2}{\rm e}^{-{\rm j} L_1 \gamma_{2,m}}
\notag\\
A_m(9,4)=-\frac{{\rm j} b_2\gamma_{2,m}}{k_2^2}{\rm e}^{{\rm j} L_1 \gamma_{2,m}}
\notag\\
A_m(9,5)=\frac{{\rm j} g {\rm e}^{-{\rm j} L_1 \gamma_{3,m}}}{k_3^2}(\alpha_m{\rm cos}\theta+\beta {\rm sin}\theta)
\notag\\
A_m(9,6)=\frac{{\rm j} g {\rm e}^{{\rm j} L_1 \gamma_{3,m}}}{k_3^2}(\alpha_m{\rm cos}\theta+\beta {\rm sin}\theta)
\notag\\
A_m(9,12)=-1,\ A_m(10,1)=\frac{{\rm j}\alpha_m}{k_1^2}{\rm e}^{-{\rm j} L_1 \gamma_{1,m}}
\notag\\
A_m(10,2)=\frac{{\rm j}\alpha_m}{k_1^2}{\rm e}^{{\rm j} L_1 \gamma_{1,m}},\ A_m(10,3)=\frac{{\rm j}\alpha_m}{k_2^2}{\rm e}^{-{\rm j} L_1 \gamma_{2,m}}
\notag\\
A_m(10,4)=\frac{{\rm j}\alpha_m}{k_2^2}{\rm e}^{{\rm j} L_1 \gamma_{2,m}}
\notag\\
A_m(10,5)=-\frac{{\rm j} {\rm e}^{-{\rm j} L_1 \gamma_{3,m}}}{k_3^2 \gamma_{3,m}}[\gamma_{3,m}^2{\rm cos}\theta+\beta(\beta {\rm cos}\theta-\alpha_m{\rm sin}\theta)] 
\notag\\
A_m(10,6)=\frac{{\rm j} {\rm e}^{{\rm j} L_1 \gamma_{3,m}}}{k_3^2 \gamma_{3,m}}[\gamma_{3,m}^2{\rm cos}\theta+\beta(\beta {\rm cos}\theta-\alpha_m{\rm sin}\theta)] 
\notag\\
A_m(10,10)=-1,\ A_m(10,12)=\frac{1}{2}{\rm j} h_2\alpha_m  
\notag\\
A_m(11,1)=\frac{{\rm j}\beta}{k_1^2}{\rm e}^{-{\rm j} L_1 \gamma_{1,m}},\ A_m(11,2)=\frac{{\rm j}\beta}{k_1^2}{\rm e}^{{\rm j} L_1 \gamma_{1,m}}
\notag\\
A_m(11,3)=\frac{{\rm j}\beta}{k_2^2}{\rm e}^{-{\rm j} L_1 \gamma_{2,m}},\ A_m(11,4)=\frac{{\rm j}\beta}{k_2^2}{\rm e}^{{\rm j} L_1 \gamma_{2,m}}
\notag\\
A_m(11,5)=-\frac{{\rm j} {\rm e}^{-{\rm j} L_1 \gamma_{3,m}}}{k_3^2 \gamma_{3,m}}[\gamma_{3,m}^2{\rm sin}\theta+\alpha_m(\alpha_m{\rm sin}\theta-\beta {\rm cos}\theta)] 
\notag\\
A_m(11,6)=\frac{{\rm j} {\rm e}^{{\rm j} L_1 \gamma_{3,m}}}{k_3^2 \gamma_{3,m}}[\gamma_{3,m}^2{\rm sin}\theta+\alpha_m(\alpha_m{\rm sin}\theta-\beta {\rm cos}\theta)] 
\notag\\
A_m(11,11)=-1,\ A_m(11,12)=\frac{1}{2}{\rm j} h_2 \beta 
\notag\\
A_m(12,1)=-\frac{2N\alpha_m\gamma_{1,m}}{k_1^2}{\rm e}^{-{\rm j} L_2 \gamma_{1,m}},\ A_m(12,2)=\frac{2N\alpha_m\gamma_{1,m}}{k_1^2}{\rm e}^{{\rm j} L_2 \gamma_{1,m}} 
\notag\\
A_m(12,3)=-\frac{2N\alpha_m\gamma_{2,m}}{k_2^2}{\rm e}^{-{\rm j} L_2 \gamma_{2,m}},\ A_m(12,4)=\frac{2N\alpha_m\gamma_{2,m}}{k_2^2}{\rm e}^{{\rm j} L_2 \gamma_{2,m}} 
\notag\\
A_m(12,5)=\frac{N {\rm e}^{-{\rm j} L_2 \gamma_{3,m}}}{k_3^2}[(\gamma_{3,m}^2+\beta^2-\alpha_m^2){\rm cos}\theta-2\alpha_m\beta {\rm sin}\theta] 
\notag\\
A_m(12,6)=\frac{N {\rm e}^{{\rm j} L_2 \gamma_{3,m}}}{k_3^2}[(\gamma_{3,m}^2+\beta^2-\alpha_m^2){\rm cos}\theta-2\alpha_m\beta {\rm sin}\theta] 
\notag\\
A_m(12,10)=-\frac{1}{2}D_{p2}(1-\nu_p)\beta^2+m_{s2}\omega^2-D_{p2}\alpha_m^2
\notag
\end{gather}
\begin{gather}
A_m(12,11)=-\frac{1}{2}D_{p2}(1+\nu_p)\alpha_m\beta 
\notag\\
A_m(13,1)=-[A+Q+b_1 E_f+2N\frac{\gamma_{1,m}^2}{k_1^2}+{\rm j} h_2 N\gamma_{1,m}\frac{\alpha_m^2+\beta^2}{k_1^2}]{\rm e}^{-{\rm j} L_2 \gamma_{1,m}} 
\notag\\
A_m(13,2)=-[A+Q+b_1 E_f+2N\frac{\gamma_{1,m}^2}{k_1^2}-{\rm j} h_2 N\gamma_{1,m}\frac{\alpha_m^2+\beta^2}{k_1^2}]{\rm e}^{{\rm j} L_2 \gamma_{1,m}} 
\notag\\
A_m(13,3)=-[A+Q+b_2 E_f+2N\frac{\gamma_{2,m}^2}{k_2^2}+{\rm j} h_2 N\gamma_{2,m}\frac{\alpha_m^2+\beta^2}{k_2^2}]{\rm e}^{-{\rm j} L_2 \gamma_{2,m}} 
\notag\\
A_m(13,4)=-[A+Q+b_2 E_f+2N\frac{\gamma_{2,m}^2}{k_2^2}-{\rm j} h_2 N\gamma_{2,m}\frac{\alpha_m^2+\beta^2}{k_2^2}]{\rm e}^{{\rm j} L_2 \gamma_{2,m}} 
\notag\\
A_m(13,5)=-(\alpha_m{\rm cos}\theta +\beta {\rm sin}\theta)\left[(\alpha_m^2+\beta^2)h_2-4{\rm j}\gamma_{3,m}-h_2\gamma_{3,m}^2\right]\frac{{\rm j} N}{2 k_3^2}{\rm e}^{-{\rm j} L_2 \gamma_{3,m}} 
\notag\\
A_m(13,6)=-(\alpha_m{\rm cos}\theta +\beta {\rm sin}\theta)\left[(\alpha_m^2+\beta^2)h_2+4{\rm j}\gamma_{3,m}-h_2\gamma_{3,m}^2\right]\frac{{\rm j} N}{2 k_3^2}{\rm e}^{{\rm j} L_2 \gamma_{3,m}} 
\notag\\
A_m(13,12)=-D_2\alpha_m^4-2D_2\alpha_m^2\beta^2-D_2\beta^4+m_{s2}\omega^2 
\notag\\
A_m(13,14)=-{\rm j}\rho_t \omega {\rm e}^{-{\rm j} L_2 \gamma_{t,m}},\ A_m(14,12)={\rm j}\omega
\notag\\
A_m(14,14)=-{\rm j}\gamma_{t,m} {\rm e}^{-{\rm j} L_3 \gamma_{t,m}} 
\notag
\end{gather}
Here, $D_{pi}$ is the in-plane stiffness, $D_i$ is the bending stiffness, $m_{si}$ is the surface density, $i=1,2$ for the incident and transmitted sides, respectively; the definition of $\theta$ is given in the \ref{app:porousexpression}; all the other elements in the matrix ${\bf A}_m$ are zero.

The non-zero element of the matrix ${\bf B}_n$ is $B_n(13,12)= K_z/l_x+ K_y \alpha_n^2/l_x$. The non-zero elements in the vector ${\bf p}$ are $p(1)={\rm j} k_z$ and $p(3)=-{\rm j} \rho_i \omega {\rm e}^{-{\rm j} h_1 k_z/2}$.


\bibliography{mybibfile}

\end{document}